\documentclass[sigconf,screen,authorversion,nonacm]{acmart}

\AtBeginDocument{%
  \providecommand\BibTeX{{%
    \normalfont B\kern-0.5em{\scshape i\kern-0.25em b}\kern-0.8em\TeX}}}

\usepackage{verbatim}
\usepackage{amsmath}
\usepackage{hyperref}

\usepackage{amssymb}
\usepackage{algorithmic}
\usepackage{algorithm}
\usepackage{makecell}

\usepackage{graphicx}
\usepackage{epstopdf}
\usepackage{array}
\usepackage{booktabs}
\usepackage{multirow}
\usepackage{caption} 
\usepackage{multicol}
\usepackage{tabularx}
\usepackage{framed}
\usepackage{enumitem}
\usepackage{threeparttable}
\usepackage{supertabular}

\graphicspath{{figs/}}

\usepackage{xcolor}

\usepackage{colortbl} 
\definecolor{jade}{rgb}{0.0, 0.66, 0.42}
\definecolor{carolinablue}{rgb}{0.6, 0.73, 0.89}
\definecolor{dkgreen}{rgb}{0,0.6,0}
\definecolor{dkblue}{rgb}{0,0.4,0.5}
\definecolor{gray}{rgb}{0.5,0.5,0.5}
\definecolor{mauve}{rgb}{0.58,0,0.82}

\definecolor{codegreen}{rgb}{0,0.6,0}
\definecolor{codegray}{rgb}{0.5,0.5,0.5}
\definecolor{codepurple}{rgb}{0.58,0,0.82}
\definecolor{codeblue}{rgb}{0,0,205}
\definecolor{backcolour}{rgb}{245,245,245}

\usepackage{listings}
\usepackage{enumitem}

\lstdefinelanguage
   [x64]{Assembler}     
   [x86masm]{Assembler} 
   {morekeywords={xend, CDQE,CQO,CMPSQ,CMPXCHG16B,JRCXZ,LODSQ,MOVSXD, %
                  POPFQ,PUSHFQ,SCASQ,STOSQ,IRETQ,RDTSCP,SWAPGS, %
                  rax,rdx,rcx,rbx,rsi,rdi,rsp,rbp, %
                  r8,r8d,r8w,r8b,r9,r9d,r9w,r9b, %
                  r10,r10d,r10w,r10b,r11,r11d,r11w,r11b, %
                  r12,r12d,r12w,r12b,r13,r13d,r13w,r13b, %
                  r14,r14d,r14w,r14b,r15,r15d,r15w,r15b}} 

\lstdefinestyle{mystyle}{
    backgroundcolor=\color{backcolour},   
    commentstyle=\color{codegreen},
    keywordstyle=\color{codeblue},
    numberstyle=\tiny\color{codegray},
    stringstyle=\color{codeblue},
    basicstyle=\ttfamily\footnotesize,
    breakatwhitespace=false,         
    breaklines=true,                 
    captionpos=b,                    
    keepspaces=true,                 
    numbers=left,                    
    numbersep=5pt,                  
    showspaces=false,                
    showstringspaces=false,
    showtabs=false,                  
    tabsize=2
}
\usepackage{url}

\usepackage{flushend}
\usepackage{balance}

\usepackage[misc]{ifsym}
\usepackage{bbding}

\usepackage{xspace}

\newcommand{\authnote}[2]{{\bf \textcolor{blue}{#1}: \em \textcolor{red}{#2}}}

\newcommand{\wh}[1]{\authnote{wh}{#1}}
\newcommand{\elsa}[1]{\authnote{elsa}{#1}}

\usepackage{xspace}

\usepackage{subfig}

 \usepackage[markup=underlined]{changes}

\theoremstyle{definition}

\newcommand{\sysname}{{\tt CEBin}\xspace}

\newcommand{\qiuhan}[1]{\textcolor{blue}{[Han: #1]}}

\newenvironment{packeditemize}{
\begin{list}{$\bullet$}{
\setlength{\labelwidth}{8pt}
\setlength{\itemsep}{0pt}
\setlength{\leftmargin}{\labelwidth}
\addtolength{\leftmargin}{\labelsep}
\setlength{\parindent}{0pt}
\setlength{\listparindent}{\parindent}
\setlength{\parsep}{0pt}
\setlength{\topsep}{3pt}}}{\end{list}}

\begin{document}

\title{\sysname:  A Cost-Effective Framework for Large-Scale Binary Code Similarity Detection}

\author{Hao Wang$^1$, Zeyu Gao$^1$, Chao Zhang$^1$, Mingyang Sun$^2$, Yuchen Zhou$^3$, Han Qiu$^1$, Xi Xiao$^4$}
\renewcommand{\authors}{Hao Wang, Zeyu Gao, Chao Zhang, Mingyang Sun, Yuchen Zhou, Han Qiu, Xi Xiao}
\affiliation{%
  \institution{$^1$Tsinghua University, Beijing, China \quad
  $^2$University of Electronic Science and Technology of China, Chengdu, China  \\
  $^3$Beijing University of Technology, Beijing, China \quad
  $^4$Tsinghua University, Shenzhen, China}
  \country{}
} 
\email{{hao-wang20,gaozy22}@mails.tsinghua.edu.cn,{chaoz,qiuhan}@tsinghua.edu.cn}

\email{2020090918021@std.uestc.edu.cn,zhouyuchen@emails.bjut.edu.cn,xiaox@sz.tsinghua.edu.cn}

\renewcommand{\shortauthors}{Wang, et al.}

\begin{abstract}
Binary code similarity detection (BCSD) is a fundamental technique for various application.
Many BCSD solutions have been proposed recently, which mostly are embedding-based, 
but have shown limited \textit{accuracy} and \textit{efficiency} especially when the volume of target binaries to search is large.
To address this issue, we propose a cost-effective BCSD framework, \sysname, which fuses embedding-based and comparison-based approaches
to significantly improve accuracy while minimizing overheads.
Specifically, \sysname utilizes a refined embedding-based approach to extract features of target code, which efficiently narrows down the scope of candidate similar code and boosts performance.
Then, it utilizes a comparison-based approach that performs a pairwise comparison on the candidates to capture more nuanced and complex relationships, which greatly improves the accuracy of similarity detection.
By bridging the gap between embedding-based and comparison-based approaches, \sysname is able to provide an effective and efficient solution for detecting similar code (including vulnerable ones) in large-scale software ecosystems. 
Experimental results on three well-known datasets demonstrate the superiority of \sysname over existing state-of-the-art (SOTA) baselines. 
To further evaluate the usefulness of BCSD in real world,
we construct a large-scale benchmark of vulnerability, offering the first precise evaluation scheme to assess BCSD methods for the 1-day vulnerability detection task. 
\sysname could identify the similar function from millions of candidate functions in just a few seconds
and achieves an impressive recall rate of $85.46\%$ on this more practical but challenging task, which are several order of magnitudes faster and $4.07\times$ better than the best SOTA baseline. Our code is available at \url{https://github.com/Hustcw/CEBin}.

\end{abstract}



\begin{abstract}
Binary code similarity detection (BCSD) is a fundamental technique for various application.
Many BCSD solutions have been proposed recently, which mostly are embedding-based, 
but have shown limited \textit{accuracy} and \textit{efficiency} especially when the volume of target binaries to search is large.
To address this issue, we propose a cost-effective BCSD framework, \sysname, which fuses embedding-based and comparison-based approaches
to significantly improve accuracy while minimizing overheads.
Specifically, \sysname utilizes a refined embedding-based approach to extract features of target code, which efficiently narrows down the scope of candidate similar code and boosts performance.
Then, it utilizes a comparison-based approach that performs a pairwise comparison on the candidates to capture more nuanced and complex relationships, which greatly improves the accuracy of similarity detection.
By bridging the gap between embedding-based and comparison-based approaches, \sysname is able to provide an effective and efficient solution for detecting similar code (including vulnerable ones) in large-scale software ecosystems. 
Experimental results on three well-known datasets demonstrate the superiority of \sysname over existing state-of-the-art (SOTA) baselines. 
To further evaluate the usefulness of BCSD in real world,
we construct a large-scale benchmark of vulnerability, offering the first precise evaluation scheme to assess BCSD methods for the 1-day vulnerability detection task. 
\sysname could identify the similar function from millions of candidate functions in just a few seconds
and achieves an impressive recall rate of $85.46\%$ on this more practical but challenging task, which are several order of magnitudes faster and $4.07\times$ better than the best SOTA baseline. Our code is available at \url{https://github.com/Hustcw/CEBin}.

\end{abstract}

\maketitle

\section{Introduction} \label{section:intro}
Binary code similarity detection (BCSD) is an emerging and challenging technique for addressing various software security problems. 
BCSD enables determining whether two binary code fragments (e.g., functions) are similar or homologous. 
BCSD can be broadly adopted for many downstream tasks like 1-day vulnerability discovery~\cite{cisco, BERTBasedBCSD, liu2018alphadiff, david2014tracelet, pewny2014leveraging, pewny2015cross, eschweiler2016discovre, david2016statistical, feng2016scalable, huang2017binsequence,feng2017extracting,david2017similarity, gao2018vulseeker, xu2017neural, david2018firmup, shirani2018binarm, vulhawk}, malware detection and classification~\cite{cesare2013control, hu2013mutantx, kim2019binary}, third-party library detection~\cite{app13010413, LibDX}, software plagiarism detection~\cite{luo2014semantics, luo2017semantics} and patch analysis~\cite{hu2016cross,xu2017spain, kargen2017towards}. BCSD's growing importance in these areas highlights its role as a versatile tool in enhancing software security.

Recently, we have witnessed numerous BCSD solutions deploying deep learning (DL) models for feature extraction and comparison~\cite{xu2017neural,liu2018alphadiff,gao2018vulseeker,redmond2018cross,zuo2018neural,ding2019asm2vec,massarelli2019safe,massarelli2019investigating,yu2020order,duan2020deepbindiff,yang2021codee}, showing that DL models can learn features of binary functions to identify similar ones across different compilers, compilation optimization levels, instruction set architectures (ISAs), or even some obfuscation techniques. 
Among them, the SOTA approaches~\cite{yu2020order, jtrans, BERTBasedBCSD, trex, palmtree, vulhawk} train \textit{large assembly language models} to learn the representation of binary code.

Despite the promising progress, current DL-based BCSD solutions are facing practical challenges when applying to real-world tasks, such as detecting 1-day vulnerabilities in the software supply scenario where the volume of target binaries to match is huge. 
For instance, once a new vulnerability is discovered in the upstream codes, efficiently and accurately identifying which downstream software has similar code and may be affected is crucial. 
For such real world tasks, a large collection of functions (e.g., all functions of the software ecosystem) must be maintained and matched against the query function (e.g., the function with the 1-day vulnerability), which brings the following three primary challenges.

First, existing BCSD methods have a poor balance between accuracy and efficiency.
Existing BCSD methods can be roughly classified into \textit {comparison-based}~\cite{dullien2005graph,10.1145/3264820.3264821, FOSSIL,liu2018alphadiff,li2019graph} and \textit {embedding-based} approaches~\cite{jtrans,palmtree,trex,yu2020order,massarelli2019safe,ding2019asm2vec,vulhawk,BERTBasedBCSD}. 
Comparison-based methods build a model to take a pair of binary functions as inputs and compare their similarity directly, which 
often have high overheads and higher accuracy.
For a given query function, it has to query the model to compare with each function in the target dataset to locate similar ones, which makes it non-scalable.  
On the other hand, embedding-based methods only take a single binary code as input and encode its higher-level features to an embedding space (i.e., numerical vectors), and then approximate the similarity of a given pair of functions in this embedding space using the vector distance (e.g., cosine), which are more scalable but have lower accuracy. 
The embedding-based approach is more efficient, since each input function only needs to be encoded once and its similar ones could be located in the embedding space with fast neighbour search algorithm. 
But the comparison-based approach in general has higher accuracy, since it takes a pair of binary functions as inputs and enables the model to learn pairwise features, while the embedding-based approach only takes one function as input and can only learn the feature of one function.


The second challenge is that existing BCSD methods cannot provide an acceptable accuracy performance (i.e., recall) when searching similar functions from a large pool of function sets. 
Pointed out in both previous study~\cite{jtrans,10.1145/3597926.3598121} and our experimental results (see Section~\ref{section:evaluation-performance}), the performance of existing BCSD declines rapidly as the scale of functions to be searched expands. 
The main reason is that \textit{the training objective of these models does not match this more challenging task}. 
For instance, existing works typically either use supervised learning to distinguish between similar or dissimilar function pairs, or employ contrastive learning to ensure the distance between similar functions is closer. 
Such models are only trained to differentiate which function \textbf{from a small number of function sets} is similar to the query function. 
This training objective cannot be simply adapted to the large-scale function datasets such as the 1-day vulnerability detection task (e.g., millions of functions to compare in the software supply chain), since in the real-world scenarios the ratio of negative samples (i.e., dissimilar functions) is way larger than the settings of model training.

The third challenge is that the community has no large-scale accessible validation dataset for BCSD tasks, such as 1-day vulnerability detection.
Existing BCSD in general only demonstrates a proof-of-concept experiment, which involves a small vulnerability dataset consisting of some CVEs (usually less than 20)~\cite{cisco,vulhawk,massarelli2019safe,jtrans,BERTBasedBCSD,LCEC-DNN,ding2019asm2vec,yang2021codee,duan2020deepbindiff,ReDeBug} and a number of target codes to search (e.g., a batch of IoT firmware). 
These methods have two drawbacks. 
(1) They choose different sets of CVEs, causing the search performance is not comparable. 
(2) They cannot evaluate the recall rate since it is impossible to determine how many vulnerabilities exist in the firmware to be tested. 
Note that, the recall rate is critical to ensure coverage comprehensiveness for understanding how 1-day vulnerabilities affect downstream software.

To address the above challenges, we propose  \sysname, a novel \underline{\texttt{C}}ost-\underline{\texttt{E}}ffective \underline{\texttt {Bin}}ary code similarity detection framework. 
\sysname fuses embedding-based and comparison-based approaches to significantly improve accuracy while minimizing overheads. 
To improve the accuracy performance of the embedding model component, \sysname proposes a Reusable Embedding Cache Mechanism (RECM) to introduce more negative samples during model fine-tuning \textit{by reusing the negative embeddings}.
This embedding model could efficiently locate similar functions with relatively high accuracy, thus greatly narrowing down the scope of candidate similar functions. 
To further improve the accuracy performance, \sysname adopts an extra comparison model component, which searches similar functions among the remained candidates in a pairwise comparison manner.

Specifically, we fuse the embedding-based and comparison-based models. 
\sysname adopts an embedding model for speed and introduces a comparison model for accuracy.
To address the inability of the comparison model to scale to large-scale functions, \sysname adopts a hierarchical approach, with the embedding model retrieving top-K functions from a large function pool, followed by the comparison model that selects the final similar ones from the top-K functions. 
With this inference process, we constrain the cost to be related to K. 
The experiments show that \sysname can increase the performance by a large margin with a high speed achieved. 

For the second challenge, we propose a Reusable Embedding Cache Mechanism (RECM) to introduce more negative samples to fine-tune the embedding model. 
As directly adding a large number of negative samples introduces significant training cost, RECM solves this challenge by \textit{maintaining an embedding cache of negative samples} during training and reusing previous embeddings. 
Then, the embedding model was trained using momentum contrastive learning~\cite{he2020momentum} by splitting the encoder model into two encoders, including (1) the query encoder to get the representation of the query function, and (2) the reference encoder to get the representation of functions in the function set. 
In this way, \sysname does not need to record the gradient of the reference encoder during training, which significantly reduces the training costs while achieving a great improvement in the embedding model's performance.

Addressing the third challenge, we aim for an objective and comprehensive evaluation of BCSD's vulnerability detection capabilities. To this end, we chose a range of widely used libraries and software incorporating 187 vulnerabilities listed in the CVE database. 
We identify the vulnerable functions corresponding to each CVE and build a benchmark with 27,081,862 functions and 12,086 vulnerable functions in total. 
With this benchmark, we take a solid step towards evaluating BCSD schemes in real-world scenarios and help future research in this domain.

We implement \sysname and evaluate it on three well-regarded BCSD datasets. 
The results show that \sysname considerably surpasses existing SOTA solutions.
In the BinaryCorp dataset, \sysname leads with an 84.5\% accuracy in identifying functions from 10,000 candidates, surpassing the current best solution's result (i.e., 57.1\%).
On two more demanding cross-architecture datasets,
\sysname attains 94.6\% and 87.0\% Recall@1, respectively, significantly outperforms the best baselines (i.e., 9.6\% and 10.9\%). 
Additionally, we conduct experiments on a large-scale cross-architecture 1-day vulnerability detection task and obtain a recall of 85.46\%, which is $4.07\times$  greater than the SOTA. 
In summary, our contributions are as follows: 

\begin{packeditemize}
    \item We propose a cost-effective BCSD framework \sysname, which fuses embedding-based and comparison-based approaches in a hierarchical inference pipeline to significantly improve accuracy performance while maintaining efficiency.

    \item We propose a Reusable Embedding Cache Mechanism (RECM) to enhance the performance of embedding models while preserving efficient training. 
    
    \item We construct a large benchmark of vulnerabilities and binary functions, offering a precise evaluation scheme to assess BCSD methods for the 1-day vulnerability detection task.

    \item We conduct thorough experiments and demonstrate the outstanding performance of \sysname for large-scale BCSD tasks, which could identify the similar function from millions in just a few seconds and achieve an impressive average recall of 85.46\%.

    \item We release \href{https://github.com/Hustcw/CEBin}{\sysname} and the large benchmark of vulnerabilities to the research community to facilitate future research.
    
\end{packeditemize}

\section{Background and Related Works}

\subsection{Binary Code Similarity Detection (BCSD)}

BCSD technique is utilized to identify the similarities between binary code fragments such as functions. 
BCSD can be adopted for many tasks like vulnerability detection, malware classification, and code plagiarism detection. 
One of the most challenging tasks is the software supply chain vulnerability detection~\cite{vulhawk,jtrans}. 
For instance, once a 1-day vulnerability is discovered in a widely-used foundational open-source component, efficiently and accurately locating the affected downstream software (mostly only binaries without source code) as comprehensive as possible is crucial. 

\begin{figure}[t!]
  \centering
  \setlength{\abovecaptionskip}{2mm}
  {
  \includegraphics[width=0.85\linewidth]{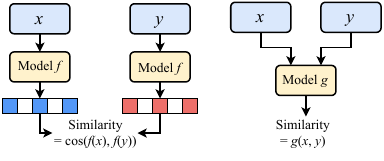}
  }
  \caption{The embedding-based model (left) represents functions $x,y$ as embeddings and calculate similarity with similarity metrics (e.g cosine). The comparison-based model (right) takes a pair of functions and outputs their similarity.}
  
  \label{fig:overview-embedding-comparison-model}
\end{figure}

Various BCSD approaches have been investigated, including graph matching~\cite{bindiff,gao2008binhunt}, tree-based methods~\cite{pewny2014leveraging}, and feature-based techniques~\cite{farhadi2014binclone,nouh2017binsign}. 
Recently, deep learning techniques have emerged as popular methods for BCSD for their accuracy and ability to learn complex features automatically.
In deep learning models, two primary methods can be distinguished: embedding-based and comparison-based approaches (see overview in Figure~\ref{fig:overview-embedding-comparison-model}).

\subsubsection{Embedding-based Approaches}

The recent development of deep neural networks (DNNs) has inspired researchers to delve into embedding-based BCSD. 
Embedding-based BCSD methods primarily focus on extracting features from functions and represent them in a lower dimension space ( i.e., ``embedding''). 
Prior research has employed DNNs as feature extractors for transforming binary functions into an embedding space. 
To determine similarity, these functions' embeddings can be compared using distance metrics. One advantage of the embedding-based approach is the use of fixed representations that can be precomputed.
When calculating the similarity between a new function and existing ones, only the new function's embedding must be extracted for distance measurement.

Genius~\cite{feng2016scalable} and Gemini~\cite{xu2017neural} employ clustering and graph neural networks (GNNs) for functional vectorization but are hampered by their reliance on control flow graph (CFG) that capture limited semantics, akin to SAFE's~\cite{massarelli2019safe} approach marred by out-of-vocabulary (OOV) challenges. Extending beyond these confines, subsequent models like GraphEmb~\cite{massarelli2019investigating} and OrderMatters~\cite{yu2020order} utilize deep neural networks to encode semantic information, with Asm2Vec~\cite{ding2019asm2vec} addressing the CFG's structural nuances through unsupervised learning. This progress sets the stage for the integration of advanced pre-trained models such as BERT, as seen in jTrans~\cite{jtrans}, Trex~\cite{trex}, and VulHawk~\cite{vulhawk}, which leverage these models' capabilities to enhance the understanding and identification of binary code functionalities.

\subsubsection{Comparison-based Approaches}

Comparison-based approaches in binary analysis directly measure function similarity using raw data or feature analysis. Methods vary: FOSSIL~\cite{FOSSIL} integrates Bayesian networks to assess free open-source software functions through syntax, semantics, and behavior. In contrast, $\alpha$-Diff~\cite{liu2018alphadiff} applies CNNs to raw bytes, requiring extensive training data. BinDNN~\cite{bindnnCNN} combines CNN, LSTM, and deep neural networks to ascertain function equivalence across compilers and architectures, while another work~\cite{shalev2018binary} decompose code into fragments for fast, accurate analysis using feed-forward neural networks. GMN~\cite{li2019graph} introduces a cross-graph attention mechanism within its DNN model for graph matching to evaluate similarity scores between graphical elements.



\subsubsection{Summary of Existing Approaches}
\label{sec:compare-embedding-analysis}
The embedding-based approach has become mainstream in BCSD research in recent years. Compared to the comparison-based approach, it can provide efficient inference and \textit{is therefore suitable for scaling to large-scale BCSD application scenarios, but has lower accuracy}. On the one hand, a recent paper~\cite{cisco} measured that the comparison-based model GMN~\cite{li2019graph} achieved the best performance among all publicly available BCSD solutions. On the other hand, we can infer this result in a theoretic way.  As shown in Figure~\ref{fig:overview-embedding-comparison-model}, given raw inputs $x$ and $y$ of two binary codes, the embedding-based model learns $f$ and uses distance metrics (e.g. cosine) to calculate similarity as $\cos(f(x),f(y))$, while the comparison-based model learns $g$ and calculates similarity as $g(x,y)$. It is obvious that, \textbf{$g(x,y)$ is more expressive than $\cos(f(x),f(y))$}. In other words, a well-trained comparison model can outperform a well-trained embedding model, to better capture the features between two binary codes. 

\subsection{Contrastive Learning}


The goal of contrastive learning is to increase the similarity between semantically similar data points, which are called embeddings, while increasing dissimilarity between semantically unrelated data points in the latent representation space. 
This is achieved by using pairwise comparison in unsupervised or self-supervised manners, measuring instance distance using a contrastive loss function.
For instance, Trex~\cite{trex} employs a pairwise loss function to minimize the distance between the ground truth. Some previous works~\cite{li2019graph,jtrans,yu2020codecmr} utilize triplet loss to reduce the distance between positive pairs and increase the distance between negative pairs. SAFE~\cite{massarelli2019safe} and OrderMatters~\cite{yu2020order} implement the output of a Siamese network as a loss function, minimizing the distance between positive pairs. Vulhawk~\cite{vulhawk} applies cross-entropy loss to reduce the distance between ground truth and maintain distance from negative pairs using a many-to-many approach.

\section{Methodology}

\subsection{Overview of the Framework} 
\begin{figure}[t!]
  \centering
  \setlength{\abovecaptionskip}{2mm}
  \includegraphics[width=0.85\linewidth]{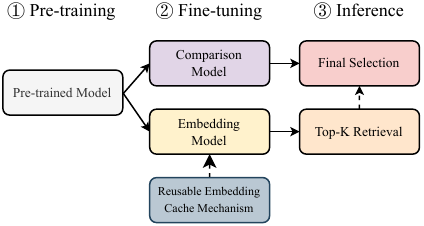}
  \caption{The Workflow of \sysname.
  }
  \label{fig:overview}
\end{figure}

The \sysname framework operates in three primary stages: pre-training, fine-tuning, and inference, depicted in Figure~\ref{fig:overview}.
In the pre-training phase, we utilize a comprehensive dataset to train a language model optimized for representing binary code.
During fine-tuning, this pre-trained language model is further refined to produce two distinct models: an embedding model and a comparison model. 
A notable enhancement during this stage is the integration of the Reusable Embedding Cache Mechanism (RECM), designed to introduce a plethora of negative samples for the fine-tuning of the embedding model.  
In the final inference phase, we employ the embedding model to retrieve the top K candidate functions closest to the query function. Subsequently, the comparison model facilitates precise final selections.

\subsection{Pre-training} 

\subsubsection{Data Preparation}
We use three datasets, BinaryCorp~\cite{jtrans}, Cisco~\cite{cisco}, and Trex~\cite{trex} as our pre-training corpus. 
We employ BinaryNinja~\footnote{\url{https://binary.ninja}} following PalmTree~\cite{palmtree} to extract functions and lift the functions to BinaryNinja's Intermediate Language (IL) to normalize binary functions across various ISAs. 
We use the WordPiece~\cite{kudo2018subword} algorithm to train a tokenizer on the whole assembly code datasets and perform a lossless encoding of assembly code without normalization on string and number, solving the problem of Out-of-Vocabulary (OOV).

\subsubsection{Model Architecture} 
We chose the Transformer~\cite{vaswani2017attention} as the base architecture for our model because both previous work~\cite{cisco} and our evaluation results in Section~\ref{section:evaluation-performance} of the baselines indicate that Transformer-based methods outperform other deep learning approaches. Because jTrans~\cite{jtrans} performs best in our evaluation, we choose to use jTrans as the base model and utilize the same pretraining tasks.

\begin{figure*}[t!]
  \centering
  \setlength{\abovecaptionskip}{2mm}
  \includegraphics[width=0.84\linewidth]{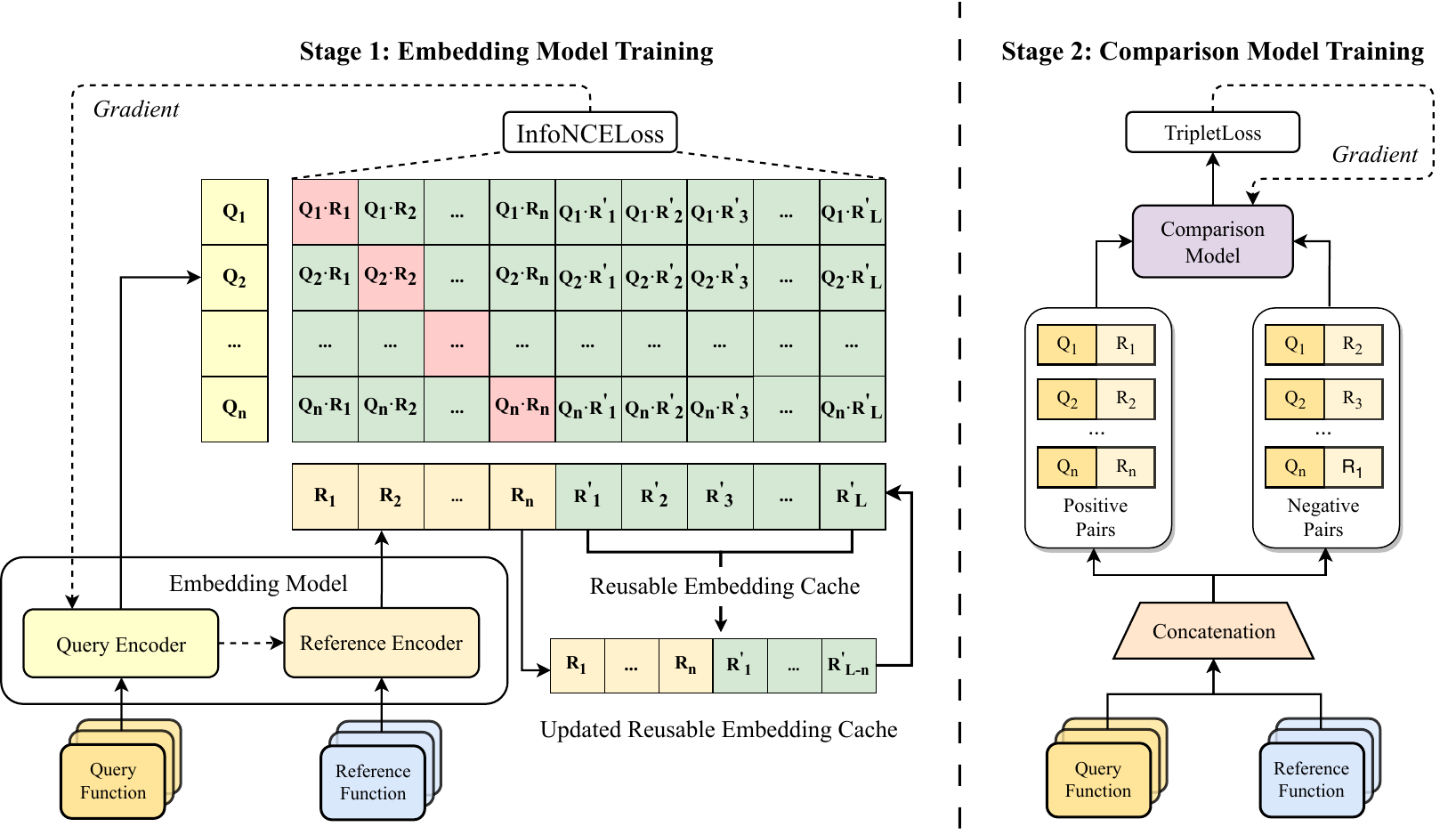}
  \caption{The illustration of fine-tuning phase for \sysname. 
  In stage 1, semantically equivalent function pairs $(Q_i, R_i)$ are encoded with query encoder and the reference encoder respectively. The corresponding pairs $(Q_i, R_i)$ are considered as positive pairs. And other pairs $(Q_i, R_j)_{i \neq j}$ along with all pairs $(Q_i, R'_j)$ containing previous reference functions in the Resuable Embedding Cache are considered as negative paris. 
  The InfoNCELoss is calcuated given positive pairs and massive negative pairs. The loss is back-propagated to update query encoder and momentum is used to update the refernece encoder. 
  In stage 2, pairs of functions are feed into model simultaneously after concatenation. $(Q_i, R_i)$ is considered as a positive pair and $(Q_i, R_{i+1})$ is considered as a negative pair. Then we use the triple loss to train the comparison model.}
  \label{fig:fine-tuning}
  \vspace{-2mm}
\end{figure*}

\subsection{Fine-tuning} 

The fine-tuning process is divided into two stages as shown in Figure~\ref{fig:fine-tuning}. Stage 1 focuses on training the embedding model, while Stage 2 trains the comparison model.

\subsubsection{RECM Integrated Embedding Model Training} 
As mentioned in the Section~\ref{section:intro}, the second challenge points out that the main issue with the current SOTA methods is that the training objective of these approaches does not match the more challenging real-world scenarios.
Note that introducing more negative samples is essential to improving the model's discrimination capabilities. 
A straightforward approach would be to directly sample a large number of negative examples during the training phase of the embedding model for contrastive learning. However, adding a large number of negative samples in an end-to-end training manner requires substantial computational resources, such as a vast amount of GPUs.

To address this challenge, we propose a Reusable Embedding Cache Mechanism (RECM) to reuse previously encoded embeddings. We first split the embedding model into a query encoder and a reference encoder. The training data is formatted as a pair $(Q_{i}, R_{i})$ fed into the model, where $Q_{i}$ and $R_{i}$ respectively represent the query function and the reference function, and they are semantically equivalent because they are compiled from the same source code. 
As shown in the stage 1 of Figure~\ref{fig:fine-tuning}, after being encoded by the embedding model, the query function and the reference function are encoded into embeddings, represented as $Q_{1:n}$ and $R_{1:n}$ respectively. We then retrieve the embeddings $R^{'}_{1:L}$ in the embedding cache, the size of the ebmedding cache is denoted as $L$. We compute the dot product between $Q_{1:n}$ and $Concate(R_{1:n}, R^{'}_{1:L})$. Only the pairs $Q_{1:n} \cdot R_{1:n}$ are positive samples, while all others are negative samples. After updating the embedding model, the embedding cache will also be updated by the newly encoded reference functions $R_{1:n}$.

While the query encoder is updated by gradients, the reference encoder is freezed during encoding and then updated using a momentum-based approach~\cite{he2020momentum}. 
When fine-tuning the embedding model, we apply the InfoNCE loss~\cite{oord2018representation} to maximize the mutual information between positive pairs and negative samples. 
The InfoNCE loss, given a positive pair $(Q_{i}, R_{i})$ and a set of negative pairs ${(Q_{i}, R_{j})}_{j \neq i}$, is defined as:
\begin{equation}
\small \mathcal{L}_{E} = - \log \frac{\exp(f(Q_{i}, R_{i}))}{\sum_{j=1}^{N} \exp(f(Q_{i}, R_{j}))}, 
\end{equation}
where $N$ is the total number of pairs, and $f(\cdot,\cdot)$ is the similarity function between two embeddings. 
We denote the parameters of query encoder and reference encoder as $\theta_{q}$ and $\theta_{r}$ respectively. 
We use momentum to update the reference encoder at the same time:
\begin{equation}
\small \theta_{r} \leftarrow m\theta_{r} + (1-m)\theta_{q}
\end{equation}
where m is the momentum coefficient and is usually set large (e.g., 0.99). During the training of the embedding model, we only update $\theta_{q}$ with back-propagation. 

With the integration of RECM, we can enlarge the size of training batches and introduce large negative samples with increasing tiny training costs. Compared to not integrating RECM, when the number of reference functions reaches $N=n+L$, training one step requires an increase of $L/n$ times in forward and backward computations, and the memory usage also increases by approximately $L/n$ times. For instance, in experiments of Section~\ref{sec:evaluation} where we used 8 V100 GPUs for training with $n=128$ and $L=8,192$. Without integrating RECM, it would require about 512 V100 GPUs which is extremely expensive. Figure~\ref{fig:BinaryCorp-Ablation-Study-K} shows the impact of the size of the embedding cache on the performance of the embedding model, which shows that our design can greatly enhance the performance of the embedding model.

\subsubsection{Comparison Model Training}
Our motivation for introducing the comparison model is inspired by image similarity detection scenarios. 
The direct comparison method enables the model to compare instances side-by-side, allowing for a token-by-token comparison of the functions. 
This approach is more precise for similarity detection tasks than the indirect comparison method.

To train the comparison model, we initialize it with the pre-trained model and modify the input to accept a pair of functions simultaneously. 
The output of the comparison model represents the similarity between the given pair of functions. 
During training, we input a batch of positive pairs, where $Q_{i}$ and $R_{i}$ form a positive pair, and $Q_{i}$ and $R_{i+1}$ serve as a negative pair. 
We concatenate the function pairs and provide them as input to the model. 
We then use the triplet loss to train the comparison model to discriminate between positive and negative pairs effectively, which can be formulated as:

\begin{equation} 
\small 
\mathcal{L}_{C} = \max(0, D(Q_{i}, R_{i}) - D(Q_{i}, R_{i+1}) + \alpha), 
\end{equation}

where $D(\cdot,\cdot)$ represents the similarity score output by comparison model and $\alpha$ is the margin of the positive and negative pairs.
By combining the fine-tuning of the embedding model in Stage 1 with the introduction of the comparison model in Stage 2, we accommodate the domain-specific requirements of the BCSD task and improve the model's ability to discern the similarity between binary code pairs effectively. 

\begin{figure*}[t!]
  \centering
  \setlength{\abovecaptionskip}{2mm}
  \includegraphics[width=0.9\linewidth]{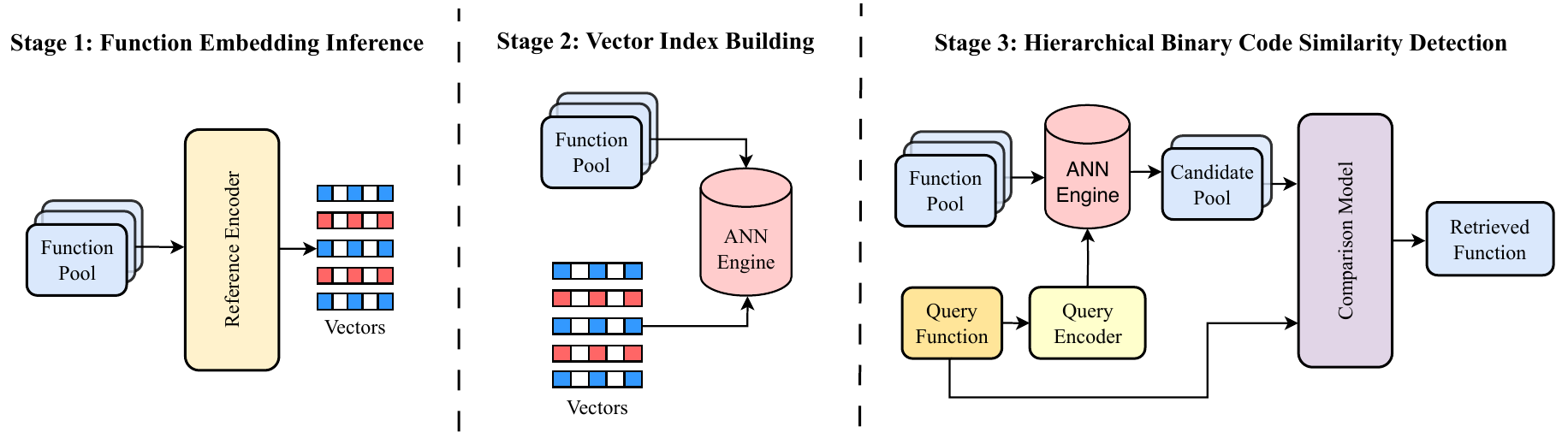}
  \caption{The illustration of inference for \sysname. In stage 1, we use a reference encoder to encode all functions we aim to compare into vectors. In stage 2, we build a vector index for each function and its corresponding vector using ANN algorithm, so that we can retrieve K most similar vectors given a query vector. In stage 3, given a query function, we use a query encoder to obtain the embedding vector and retrieve the top-K closest functions from the function pool using a pre-built vector index. Then the K candidate functions along with the query function are fed into the comparison model to perform the final selection.}
  \label{fig:inference-workflow}
\end{figure*}

\subsection{Inference} 
The \sysname inference process has three stages as shown in Figure~\ref{fig:inference-workflow}. 
The three stages perform function embedding inference, vector index building, and binary code similarity detection, respectively. This hierarchical design aims to balance performance and inference cost, integrating the advantages of both the embedding model and the comparison model.

\textbf{Stage 1: Function Embedding Inference.} In the first stage, we use the  embedding model to construct embedding vectors for each function within the function pool that we aim to compare. As shown in Figure~\ref{fig:inference-workflow}, we employ the reference encoder from the embedding model to generate vectors for the functions in the function pool.

\textbf{Stage 2: Vector Index Building.} In the second stage, we build a vector index for each function in the function pool using the constructed embedding vectors. 
We use the approximate nearest neighbor (ANN) algorithm to enable efficient inference. 
ANN approximates the nearest neighbors in high-dimensional spaces, allowing for fast similarity search and comparison among the function pool's embedding vectors. 
By building a vector index using ANN, our model can handle large-scale BCSD scenarios in a resource-efficient manner.

\textbf{Stage 3: Hierarchical Binary Code Similarity Detection.} The third stage of the inference process involves utilizing the fine-tuned embedding model and the comparison model from \sysname to perform BCSD. 
Given a query function, we first use the embedding model to retrieve the top-K closest functions from the function pool using the ANN-based vector index. 
This helps to narrow down the most similar functions while maintaining high efficiency. 
After obtaining the k candidate functions, we use the comparison model to do BCSD with the query function and candidate functions.

The hierarchical combination of the embedding model and the comparison model ensures that our BCSD method is both efficient and accurate.  
The computational cost of using the comparison model by concatenating the query function and top-K candidate functions as input is manageable and doesn't increase substantially as the size of the original function pool grows. 
This allows \sysname to effectively identify the most similar binary code sequences within massive functions.

Our extensive experiments in Section~\ref{sec:evaluation} show that \sysname's approach in conjunction with the ANN engine is highly cost-effective, supporting large-scale comparisons with high efficiency. 
Furthermore, the comparison model significantly enhances the detection results by providing a fine-grained similarity assessment. We achieve better accuracy without incurring a dramatic increase in computational cost through the hierarchical inference framework. 
\section{Experimental Setup}
We compare \sysname against multiple baselines: Genius~\cite{feng2016scalable}, Gemini~\cite{xu2017neural}, SAFE~\cite{massarelli2019safe}, Asm2Vec~\cite{ding2019asm2vec}, GraphEmb~\cite{massarelli2019investigating}, OrderMatters~\cite{yu2020order}, Trex~\cite{pei2022learning}, GNN and GMN~\cite{li2019graph}, and jTrans~\cite{jtrans}. We implement \sysname and baselines using Faiss~\cite{johnson2019billion} and Pytorch~\cite{pytorch}. Our experiments are conducted on several servers to accelerate training. The GPU setup includes 8 Nvidia-V100. 
The experiment environment consists of three Linux servers running Ubuntu 20.04 with Intel Xeon 96-core and equipped with 768GB o f RAM.

We evaluate \sysname with the following three datasets. We label functions as equal if they share the same name and were compiled from the same source code.

\begin{packeditemize}
\item \textbf{BinaryCorp}~\cite{jtrans} is sourced from the ArchLinux official repositories and Arch User Repository. 
Compiled using GCC 11.0 on X64 with various optimizations, it features a highly diverse set of projects.

\item \textbf{Cisco Dataset}~\cite{cisco} comprises seven popular open-source projects, it yields 24 distinct libraries upon compilation. 
Binaries in the Cisco dataset are compiled with GCC and Clang compilers, spanning four versions each, across six ISAs (x86, x64, ARM32, ARM64, MIPS32, MIPS64) and five optimization levels (O0-O3, Os). This setup allows for cross-architecture analysis and evaluation of compiler versions, with a moderate number of projects.

\item \textbf{Trex Dataset}~\cite{pei2022learning} is built upon binaries released by~\cite{pei2022learning}, which consists of ten libraries chosen to avoid overlap with the Cisco dataset. Similar to the Cisco Dataset, the Trex dataset facilitates cross-architecture and cross-optimization evaluation.
\end{packeditemize}



\begin{table*}[t!]
\caption{Comparison between \sysname and baselines for the cross-optimization task on \texttt{BinaryCorp}-3M (Poolsize=10,000)}
\centering
\scalebox{0.8}{
\begin{tabular}{c|ccccccc|ccccccc}
\Xhline{1pt}
\hline
             & \multicolumn{7}{c|}{\textbf{MRR}}                             & \multicolumn{7}{c}{\textbf{Recall@1}}                               \\ \hline
\textbf{Models}       & \textbf{O0,O3} & \textbf{O1,O3} & \textbf{O2,O3} & \textbf{O0,Os} & \textbf{O1,Os} & \textbf{O2,Os} & \textbf{Average} & \textbf{O0,O3} & \textbf{O1,O3} & \textbf{O2,O3} & \textbf{O0,Os} & \textbf{O1,Os} & \textbf{O2,Os} & \textbf{Average}  \\ \hline
Genius &0.041   &0.193  &0.596  &0.049  &0.186  &0.224    &0.214  &0.028  &0.153  &0.538  &0.032  &0.146  &0.180  & 0.179   \\ 
Gemini         & 0.037 & 0.161 & 0.416 & 0.049 & 0.133 & 0.195 & 0.165   & 0.024 & 0.122 & 0.367 & 0.030 & 0.099 & 0.151 & 0.132    \\
GNN & 0.048 & 0.197 & 0.643 & 0.061 & 0.187 & 0.214 & 0.225 & 0.036 &  0.155 & 0.592 & 0.041 & 0.146 & 0.175 & 0.191 \\ 
GraphEmb     & 0.087 & 0.217 & 0.486 & 0.110 & 0.195 & 0.222 & 0.219 & 0.050 & 0.154 & 0.447 & 0.063 & 0.135 & 0.166 & 0.169          \\
OrderMatters & 0.062 & 0.319 & 0.600 & 0.075 & 0.260 & 0.233 & 0.263 & 0.040 & 0.248 & 0.535 & 0.040 & 0.178 & 0.158 & 0.200          \\ 
SAFE         & 0.127 & 0.345 & 0.643 & 0.147 & 0.321 & 0.377 & 0.320 & 0.068 & 0.247 & 0.575 & 0.079 & 0.221 & 0.283 & 0.246  \\
Asm2Vec      & 0.072 & 0.449 & 0.669 & 0.083 & 0.409 & 0.510 & 0.366 & 0.046 & 0.367 & 0.589 & 0.052 & 0.332 & 0.426 & 0.302  \\
Trex & 0.118 & 0.477 & 0.731 & 0.148 & 0.511 & 0.513 & 0.416 & 0.073 & 0.388 & 0.665 & 0.088 & 0.422 & 0.436 & 0.345 \\ 
jTrans       & 0.475 & 0.663 & 0.731 & 0.539 & 0.665 & 0.664 & 0.623 & 0.376 & 0.580 & 0.661 & 0.443 & 0.586 & 0.585 & 0.571 \\ \hline
\sysname-E & 0.787 & 0.874 & 0.924 & 0.858 & 0.909 & 0.893 & 0.874 & 0.710 & 0.818 & 0.885 & 0.795 & 0.863 & \textbf{0.842} & 0.819 \\
\sysname & \textbf{0.850} & \textbf{0.886} & \textbf{0.953} & \textbf{0.903} & \textbf{0.927} & \textbf{0.895} & \textbf{0.902} & \textbf{0.776} & \textbf{0.826} & \textbf{0.920} & \textbf{0.839} & \textbf{0.874} & 0.834 & \textbf{0.845} \\
\Xhline{1pt}
\end{tabular}
}\label{tab:binarycorp-10000}
\end{table*}





Many previous work~\cite{cisco,massarelli2019safe,vulhawk,pei2022learning,xu2017neural} use the area under curve (AUC) of the receiver operating characteristic (ROC) curve or precision to evaluate the performance of BCSD solutions.
But this metric is too simple for existing solutions so SOTA BCSD solutions perform similarly. 
However, we notice that previous works~\cite{jtrans,cisco} announced that ranking metrics, the mean reciprocal rank (MRR), and the recall (Recall@K) are more practical for BCSD especially when the size of the function pool becomes very large. 
Therefore, we use MRR and Recall@1 following~\cite{jtrans} to evaluate and compare the performance of \sysname and the baseline methods.

\iftrue

During the pre-training stage, we use a more extensive dataset and adopt the same configuration of jTrans~\cite{jtrans}. The pre-training model is optimized with Adam~\cite{adam} with parameters of $\beta_{1}=0.9, \beta_{2}=0.98, \epsilon = 1e\text{-}12$, and an $L_{2}$ weight decay of 0.01. 
The model trains on mini-batches consisting of $B=128$ binary functions with a cap of $T=512$ tokens per function.

During the fine-tuning phase, we assign the temperature $T=0.05$, the negative queue size $L = 8192$, and the momentum $m = 0.99$. The comparison model utilizes a margin of $\alpha = 0.25$ for Triplet Loss. 
The models are trained using the Adam algorithm~\cite{adam} with these parameters: $\beta_{1}=0.9, \beta_{2}=0.999, \epsilon = 1e\text{-}8$, and an $L_{2}$ weight decay of 0.0001. 
\fi

For \sysname's inference phase, we retrieve the top-50 closest function for experiments on BinaryCorp, Cisco, and Trex datasets. We choose K=50 because we find that the embedding model's Recall@50 is almost close to 1.0 as shown in the Section~\ref{section:recs}, thus providing a sufficiently good set of candidate functions for the comparison model. We retrieve top-300 closest function for  the vulnerability search experiments because the maximum number of vulnerable functions could be up to 240 for each query.

\section{Evaluation}\label{sec:evaluation}

To prove \sysname's effectiveness in addressing previous challenges, we propose these research questions (RQs):
\begin{packeditemize}
    \item \textbf{RQ1:} How does \sysname perform compared to SOTA BCSD solutions in different settings, including cross-architecture, cross-compilers, and cross-optimizations?
    \item \textbf{RQ2:} How do the design choices within the \sysname framework contribute to the overall performance?
    \item \textbf{RQ3:} How does \sysname perform in vulnerability search over a challenging vulnerability searching benchmark?
    \item \textbf{RQ4:} How is the generalization ability of the \sysname?
    \item \textbf{RQ5:} What is the inference time cost of \sysname compared with other SOTA baselines?
\end{packeditemize}


\begin{table*}[t!]
\centering
\caption{Results of different binary similarity detection approaches on Cisco (poolsize=10,000)}
\scalebox{0.8}{
\begin{tabular}{c|cccccc|cccccc}
\Xhline{1pt}
 & \multicolumn{6}{c|}{\textbf{MRR}} & \multicolumn{6}{c}{\textbf{Recall@1}} \\ \hline
\textbf{Models} & \textbf{XA} & \textbf{XC} & \textbf{XO} & \textbf{XA+XO} & \textbf{XC+XO} & \makecell[c]{\textbf{XA+}\\\textbf{XC+XO}} & \textbf{XA} & \textbf{XC} & \textbf{XO} & \textbf{XA+XO} & \textbf{XC+XO} & \makecell[c]{\textbf{XA+}\\\textbf{XC+XO}} \\ \hline
GNN & 0.205 & 0.158 & 0.104 & 0.119 & 0.189 & 0.093 & 0.129 & 0.104 & 0.080 & 0.084 & 0.165 & 0.063 \\ 
Trex & 0.085 & 0.401 & 0.410 & 0.145 & 0.313 & 0.124 & 0.052 & 0.341 & 0.360 & 0.113 & 0.268 & 0.096 \\ \hline
\sysname-E & 0.760 & 0.907 & 0.859 & 0.817 & 0.866 & 0.766 & 0.692 & 0.871 & 0.816 & 0.766 & 0.823 & 0.706 \\
\sysname & \textbf{0.977} & \textbf{0.992} & \textbf{0.973} & \textbf{0.978} & \textbf{0.984} & \textbf{0.961} & \textbf{0.968} & \textbf{0.988} & \textbf{0.963} & \textbf{0.969} & \textbf{0.977} & \textbf{0.946} \\ \Xhline{1pt}
\end{tabular}
}\label{tab:Cisco-Poolsize-10k}
\end{table*}

\begin{table}[t!]
\centering
\caption{Results of different binary similarity detection approaches on Trex (poolsize=10,000)}
\scalebox{0.8}{
\begin{tabular}{c|ccc|ccc}
\Xhline{1pt}
 & \multicolumn{3}{c|}{\textbf{MRR}} & \multicolumn{3}{c}{\textbf{Recall@1}} \\ \hline
\textbf{Models} & \textbf{XA} & \textbf{XO} & \textbf{XA+XO} & \textbf{XA} & \textbf{XO} & \textbf{XA+XO} \\ \hline
Trex & 0.142 & 0.218 & 0.175 & 0.065 & 0.123 & 0.107 \\ 
GNN & 0.163 & 0.148 & 0.151 & 0.145 & 0.102 & 0.109 \\ \hline
\sysname-E & 0.612 & 0.646 & 0.576 & 0.509 & 0.553 & 0.474 \\
\sysname & \textbf{0.911} & \textbf{0.933} & \textbf{0.911} & \textbf{0.882} & \textbf{0.906} & \textbf{0.870} \\ \Xhline{1pt}
\end{tabular}
}\label{tab:Trex-Poolsize-10000}
\end{table}

\subsection{Performance~(RQ1)}\label{section:evaluation-performance}

\subsubsection{Cross-Optimizations: BinaryCorp}
In this experiment, we assess \sysname's performance on the BinaryCorp dataset, which includes x64 binaries compiled with \texttt{GCC-11} across various optimization levels (O0, O1, O2, O3, and Os). 
We conduct extensive experiments to evaluate the performance of the selected baselines, which are limited to a single architecture and those supporting cross-architecture. 
We evaluate the performance of cross-optimization BCSD tasks with varying difficulty optimization pairs (e.g., O0 v.s. O3) while maintaining consistent experimental setups with previous work~\cite{jtrans} for fair comparison. 
We report the experimental results for function poolsize 10,000 as shown in Tables~\ref{tab:binarycorp-10000}. 
\sysname-E denotes for the embedding model of \sysname.

The experimental results in Table~\ref{tab:binarycorp-10000} demonstrate that \sysname significantly outperforms all baselines. 
\sysname outperforms the best-performing baseline jTrans by significantly improving MRR by 44.8\% and Recall@1 by 47.9\%. 
The experimental results show the advantage of \sysname in cross-optimization tasks, improving the effectiveness of the embedding model training by using more negative samples during training.

\subsubsection{Cross-Architectures, Compilers, and Optimizations: Cisco and Trex Dataset}
We evaluate \sysname and baselines on Cisco and Trex datasets across various factors, including architectures, compilers, optimization options, and their combinations. 
In this experiment, we select several cross-architecture baselines for comparisons, such as GNN, and Trex. 
Consistent with previous work, we train \sysname and GNN on Cisco's training set and assess performance on Cisco's test set. 
As previous research~\cite{cisco} highlights, retraining Trex on Cisco dataset is challenging, we directly use the model released by Trex authors.

To ensure comprehensive testing, we employ six different evaluation tasks. 
(1) XO refers to function pairs with varying optimizations but identical compiler, compiler version, and architecture. 
(2) XC refers to function pairs with different compilers but the same architecture and optimization. 
(3) XA refers to function pairs with varying architectures but identical compiler, compiler version, and optimization. 
(4) XC+XO refers to function pairs with different compilers and optimizations but the same architecture. 
(5) XA+XO refers to function pairs with varying architectures and optimizations but identical compiler and compiler versions. 
(6) XA+XC+XO refers to function pairs from any architecture, compiler, compiler version, and optimization. 
We test the six tasks on the Cisco Dataset. 
We test tasks (1), (3), and (5) on the Trex dataset since it only uses one compiler (i.e., GCC 7.5). 
We evaluate performance in a much more challenging scenarios where poolsize=10,000 compared to previous works.

Table~\ref{tab:Cisco-Poolsize-10k}--\ref{tab:Trex-Poolsize-10000} report the experimental results. 
The results reveal that \sysname significantly outperforms baselines in mixed cross-architecture, cross-compiler, and cross-optimization tasks. 
Compared to the best baseline Trex, the MRR increases from 0.124 to 0.961, and Recall@1 increases from 0.096 to 0.946.
On the Trex Dataset, 
\sysname outperforms the best result of the baseline, with MRR increasing from 0.175 to 0.911 and Recall@1 increasing from 0.109 to 0.870.

The results demonstrate \sysname's advantages in challenging BCSD tasks. 
Training more efficiently on a larger quantity of negative samples enables the embedding model to perform better.
As our training goal involves discriminating similar binary codes in larger batches of negatives, \sysname significantly outperforms the baseline, especially for this challenging large poolsize settings. 
Additionally, integrating the comparison model further enhances performance as it achieves more fine-grained similarity detection. 
In the XA+XO experiment conducted on the Trex dataset, the comparison model significantly improves Recall@1 from 0.474 to 0.870. 

\begin{figure}[t!]
  \centering
  \setlength{\abovecaptionskip}{2mm}
  \scalebox{0.95}{
  \includegraphics[width=0.97\linewidth]{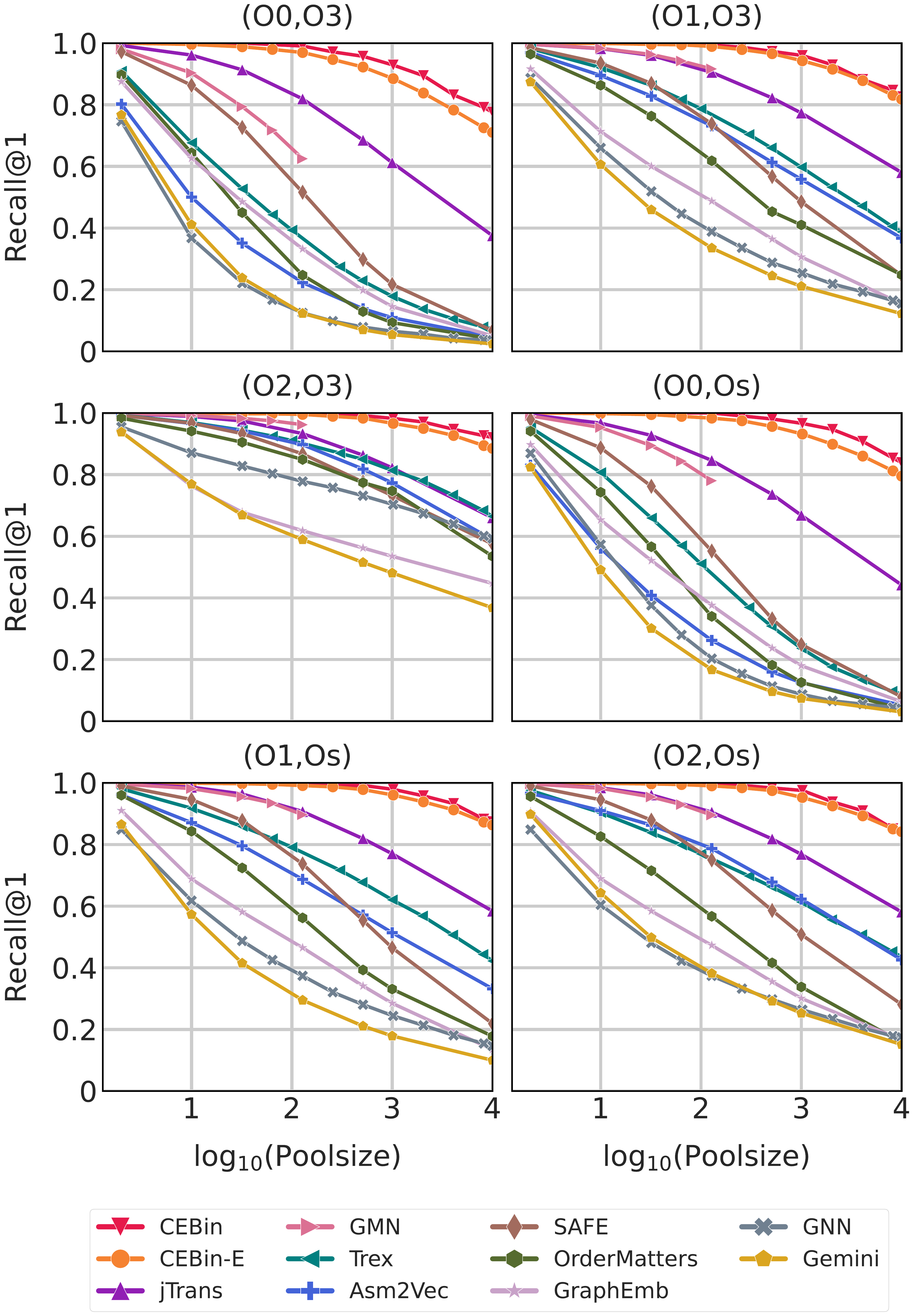}
  }
  \caption{The performance of different binary similarity detection methods on BinaryCorp. The x-axis is logarithmic and denotes the poolsize.}
  
  \label{fig:BinaryCorp-Recall@1-Poolsize}
  \vspace{-2mm}
\end{figure}

\begin{figure}[t!]
  \centering
  \setlength{\abovecaptionskip}{2mm}
  \scalebox{0.95}{
  \includegraphics[width=0.97\linewidth]{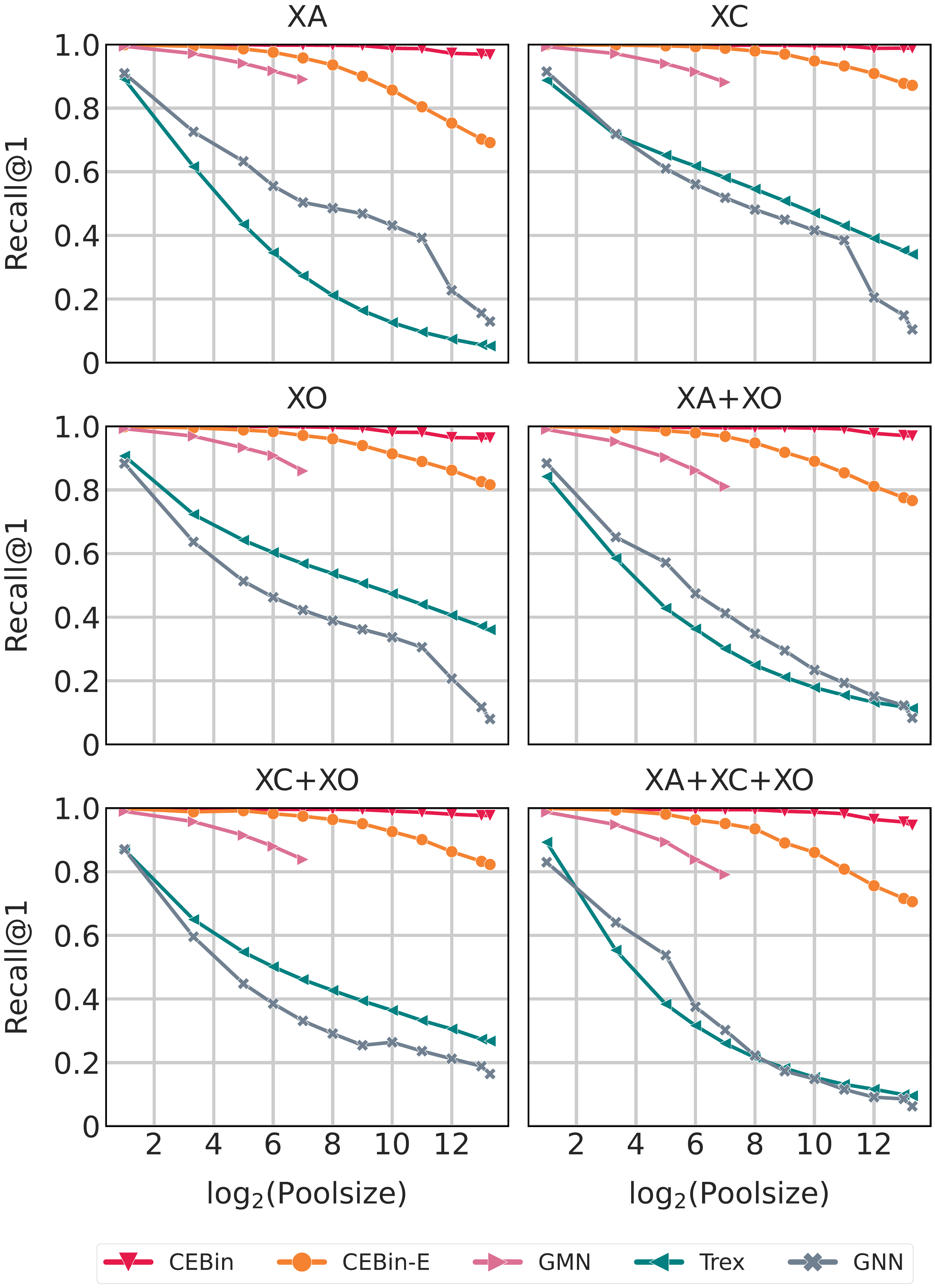}
  }
  \caption{ The performance of different binary similarity detection methods on Cisco Dataset. The x-axis is logarithmic and denotes for the poolsize.}
  \label{fig:Cisco-Recall@1-Poolsize}
  \vspace{-1mm}
\end{figure}

\begin{figure}[t!]
  \centering
  \setlength{\abovecaptionskip}{2mm}
  \scalebox{0.9}{
  \includegraphics[width=0.97\linewidth]{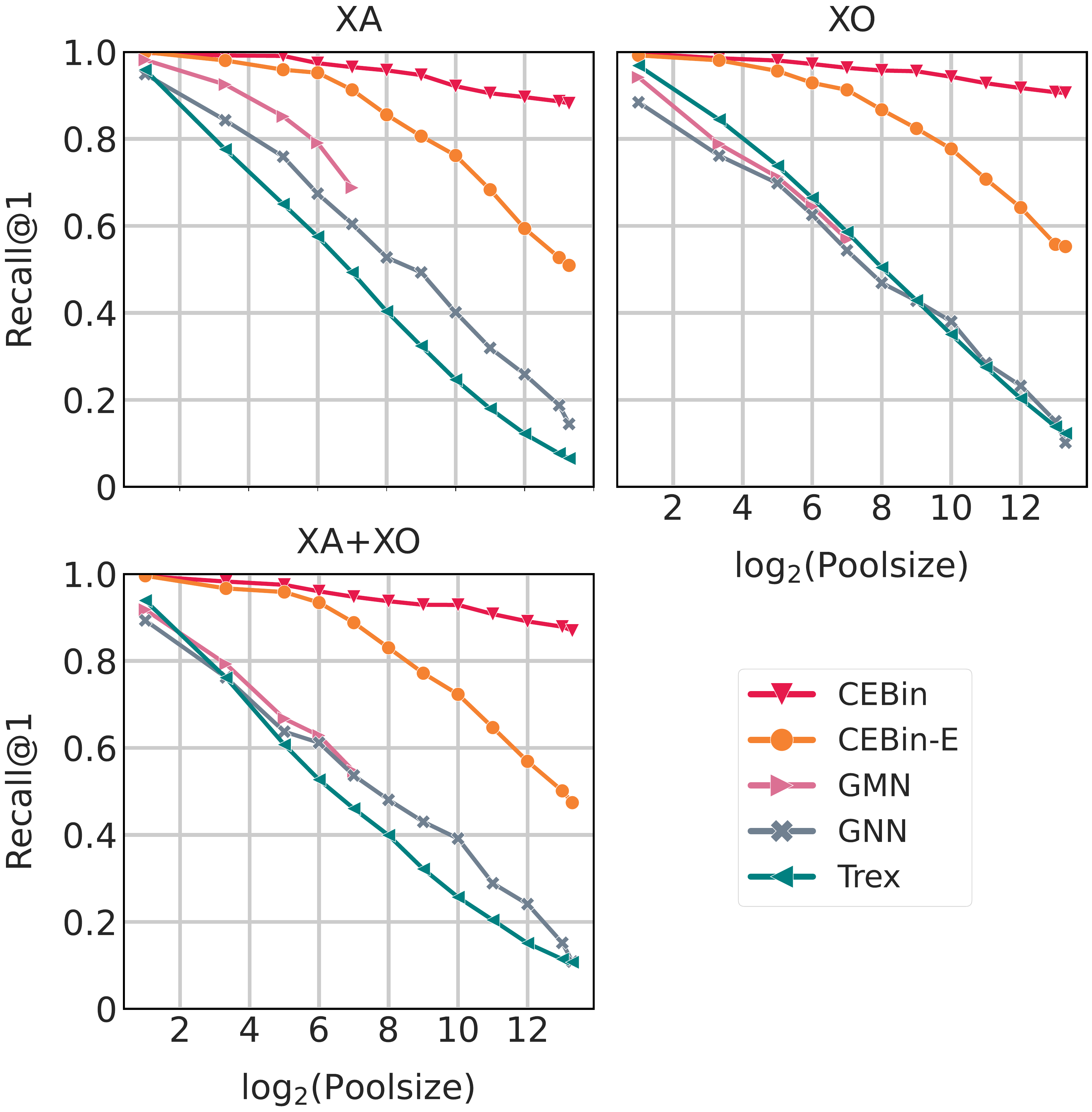}
  }
  \caption{ The performance of different binary similarity detection methods on Trex Dataset. The x-axis is logarithmic and denotes for the poolsize.}
  
  \label{fig:Trex-Recall@1-Poolsize}
\end{figure}

\subsubsection{The Impact of Poolsize}
As indicated in Section~\ref{section:intro}, for practical tasks like 1-day vulnerability detection in software supply chains, maintenance of a particularly large poolsize is necessary and valuable. 
However, in our prior experiments, we discover that as poolsize increases, performance declines across the three datasets. 
Thus, we explore the influence of different poolsize while maintaining other settings. 
The poolsize is set as $2^i, i \in [1, 13]$, and 10,000.
We record Recall@1 for different poolsize. 

The Figure~\ref{fig:BinaryCorp-Recall@1-Poolsize}--\ref{fig:Trex-Recall@1-Poolsize} presents the results, clearly showing that as the poolsize increases, the relative performance of all baselines is inferior to \sysname. 
Furthermore, the decline in the performance of \sysname is not so obvious which suggests that \sysname is more capable of addressing large poolsize settings. 
The results also show that \sysname offers a greater performance enhancement compared to \sysname-E in more difficult scenarios such as O0 and O3 optimization options in the BinaryCorp experiment and the XA+XC+XO in the Cisco dataset, where binary functions exhibit larger discrepancies. 
\sysname, trained on the Cisco dataset, displays remarkable poolsize robustness on the Trex dataset, demonstrating outstanding generalization performance. 
Finally, we emphasize that when the poolsize is small (e.g., poolsize=2), the difference in recall@1 among different methods is tiny, indicating results measured with a very small poolsize in many existing works cannot accurately represent these BCSD solutions' performance in real-world.

\subsection{Impact of Our Design Choices ~(RQ2)}\label{section:evaluation-ablation-study}
In this section, we aim to validate the effects of our two core designs: introducing more negative samples through RECM during training and hierarchical inference on performance.

\begin{figure}[t!]
  \centering
  \setlength{\abovecaptionskip}{2mm}
  \scalebox{0.95}{
  \includegraphics[width=1\linewidth]{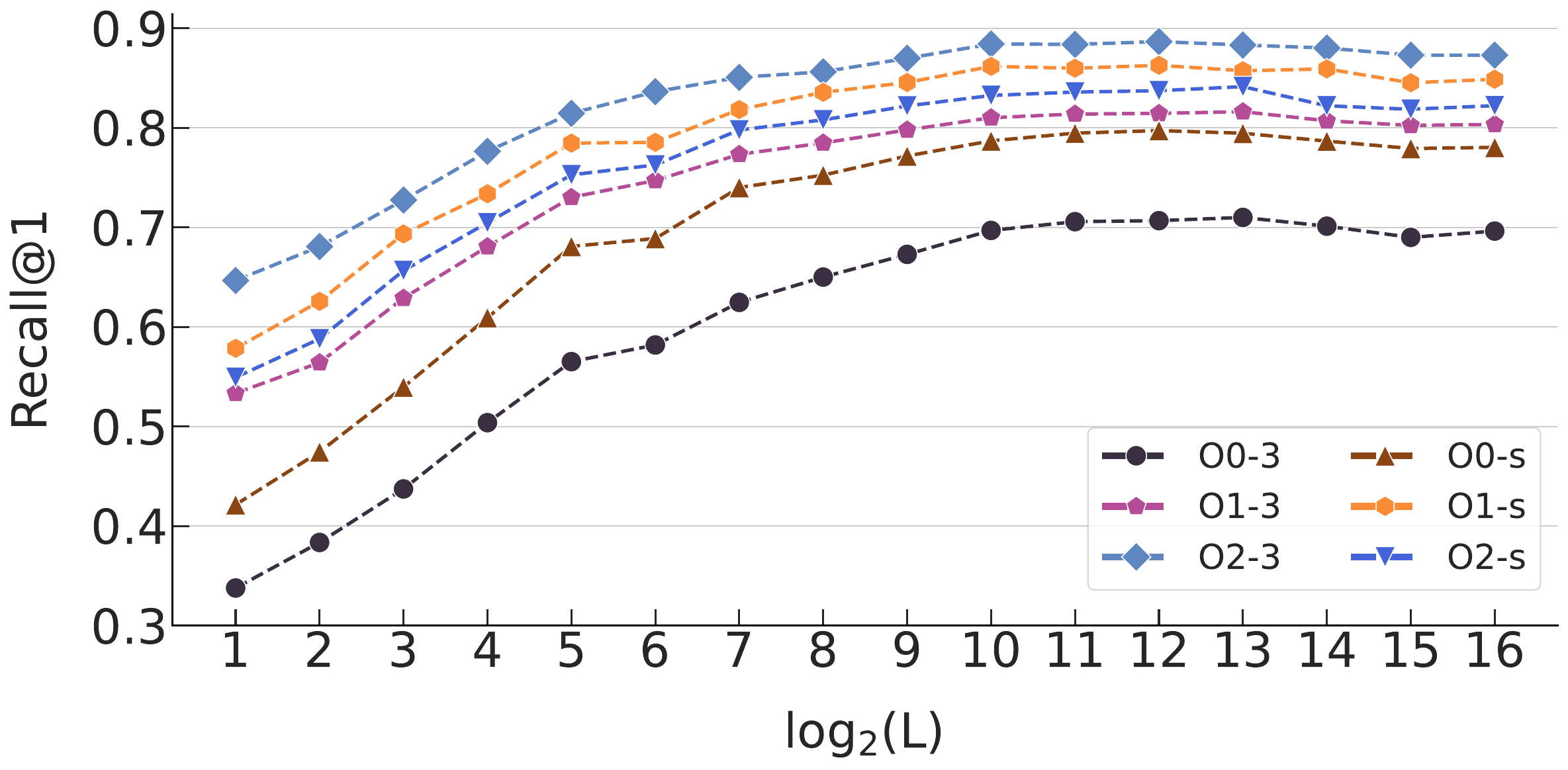}
  }
  \caption{ The performance of \sysname-E on BinayCorp using different size of embedding cache.}
  
  \label{fig:BinaryCorp-Ablation-Study-K}
\end{figure}

\subsubsection{Reusable Embedding Cache Mechanism}\label{section:recs}
To investigate the impact of the number of negative samples during training, we keep other parts of the embedding model training consistent and only change the size of the RECM, also represents the number of negative samples, used during training. 
We set L, the size of RECM, as powers of 2 ranging from 2 to 65536. Then we evaluate the Recall@1 for different optimization pairs at a poolsize of 10,000 on BinaryCorp. 
The experimental results are displayed in Figure~\ref{fig:BinaryCorp-Ablation-Study-K}, where the x-axis represents the poolsize (a logarithmic axis).

Based on the experimental results, we observe that an increased number of negative samples substantially improves the overall effectiveness of the embedding model across various cross-optimization tasks. 
For example, in the most challenging task (O0 and O3), Recall@1 increases from 0.337~(L=2) to 0.709~(L=8192). 
In less challenging scenario such as comparing O2 and O3, recall@1 rose from 0.647~(L=2) to 0.887~(L=4096). 
Interestingly, we found that a larger L does not always lead to better results. 
This can be attributed to the continuous update of encoders during the encoding process where RECM employed for training. 
Although a large momentum maintains a slow update to ensure consistency of embeddings in the embedding cache, excessive reuse with an exceedingly large size L introduces inconsistencies that slightly reduce performance. 
According to experimental results, performance improvement begins to dwindle when L exceeds 1024. 
A comparative analysis revealed that the optimal average performance among the six cases is attained at L=8,192 with an average recall@1 of 0.819. 
The experimental results verify that the integration of RECM significantly improves the performance.

\begin{figure}[t!]
  \centering
  \setlength{\abovecaptionskip}{2mm}
  \scalebox{0.95}{
  \includegraphics[width=.9\linewidth]{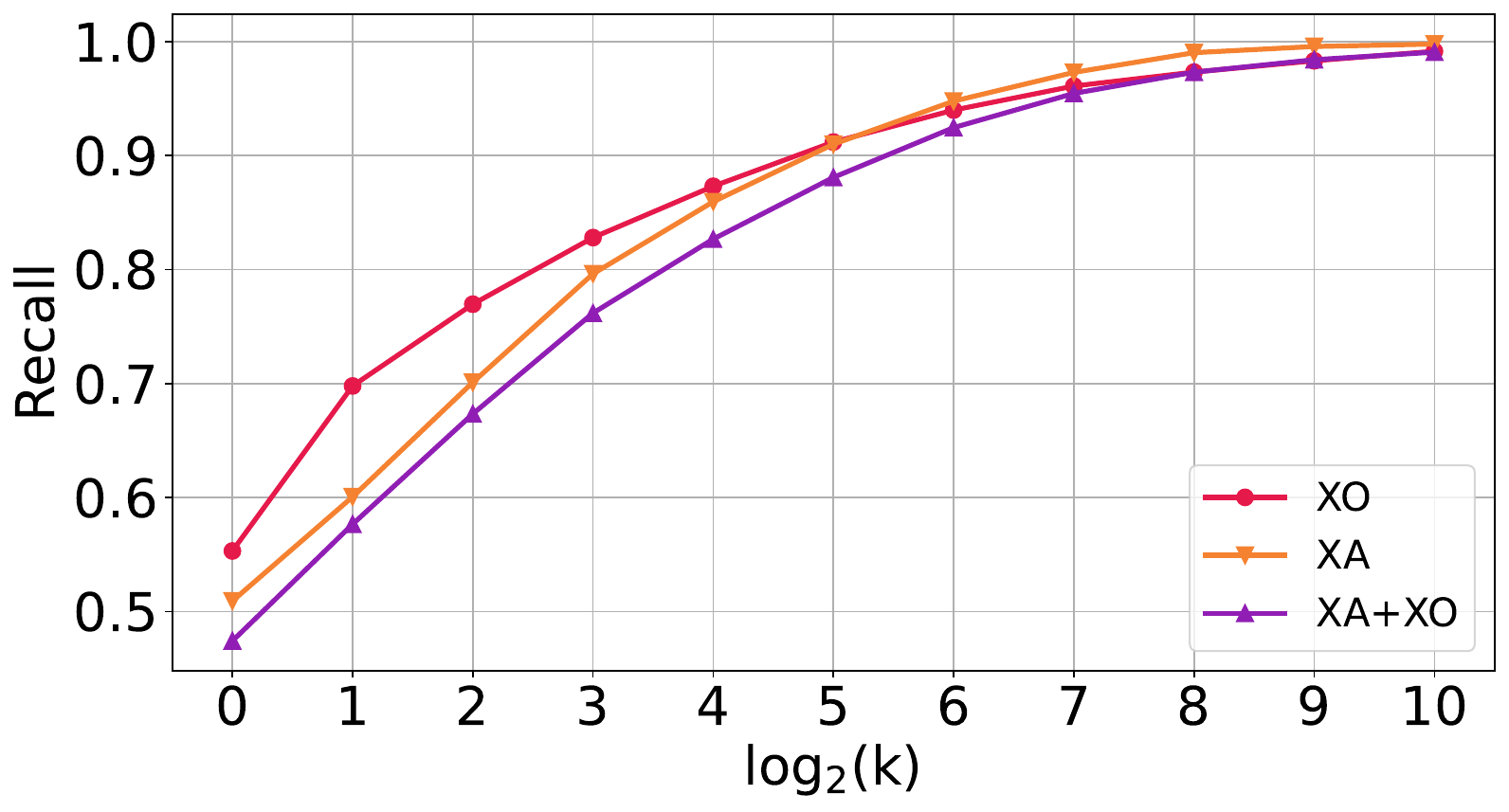}
  }
  \caption{Recall@K of \sysname-E on Trex Dataset for poolsize=10,000.}
  \label{fig:Recall@K-Trex}
\end{figure}

\subsubsection{Comparison Model}
To investigate the role of the comparison model in \sysname, we examine both \sysname and \sysname-E across all RQ1 experiments, where \sysname-E employs only the embedding model. 
The enhancement of \sysname in relation to \sysname-E indicates the point of the comparison model during hierarchical inference. 
Results are shown in Tables~\ref{tab:binarycorp-10000}--\ref{tab:Trex-Poolsize-10000} and Figures~\ref{fig:BinaryCorp-Recall@1-Poolsize}--\ref{fig:Trex-Recall@1-Poolsize}.

Our findings reveal that the comparison model delivers performance gains in cross-architecture, cross-compiler, and cross-optimization contexts, with larger improvements observed in more challenging tasks. 
In the cross-optimization task of the BinaryCorp dataset with poolsize=10,000, the comparison model boosts the average Recall@1 by 3.0\%, with the most arduous O0, O3 task elevating Recall@1 by 9.3\%. 
For the XA+XC+XO task in the Cisco dataset when poolsize=10,000, \sysname's Recall@1 rises by 34.0\% compared to \sysname-E. 
In the XA+XO task on the Trex dataset when poolsize=10000, Recall@1 increases from 0.474 to 0.870, which is an 83.5\% improvement.
As \sysname is trained on Cisco and the training set lacks the GCC7.5 compiler, the Trex dataset represents an out-of-distribution (OOD) dataset. 
The performance significantly declines with only the embedding model, yet incorporating the comparison model markedly heightens \sysname's robustness. 

To show the potential improvement the comparison-based model can bring, 
Figure~\ref{fig:Recall@K-Trex} presents the Recall@K of \sysname-E on the Trex dataset. The experimental result shows that the Recall@50 of \sysname-E is significantly higher than Recall@1, which indicates the potential improvements that can be brought about by using the comparison model. However, the Recall@50 of \sysname-E is only slightly higher than the Recall@1 of \sysname, which suggests that our comparison model performs very well. The experiments demonstrate that the comparison model effectively learns more intricate features, augmenting BCSD performance. It's worth noting that, the Recall@50 of the embedding model is almost convergent. Considering the balance between overhead and performance, we chose K=50 for all the experiments presented earlier as the candidate results are good enough for comparison model.

\subsection{Vulnerability Detection~(RQ3)}\label{section:evaluation-vulnerability-detection} 
To develop a realistic vulnerability dataset, we gather commits that fix 187 CVEs from 5 projects, including \texttt{curl}, \texttt{vim}, \texttt{libpng}, \texttt{openssh} and \texttt{openssh-portable}. 
These projects are compiled using multiple compilers and architectures to represent the diverse configurations found in real-world software supply chains. 
We provides details about the dataset including the number of vulnerable functions and the poolsize corresponding to each CVE search in our released code.
We first pinpoint relevant functions within commits addressing CVEs by analyzing their root causes, given the CVE and its associated commit. 
Next, we evaluate \sysname by establishing a search pool comprising all compiled functions. 
We then select one vulnerable function to serve as a query and attempt to identify other instances of the same vulnerability in the entire pool. 
Given $m$ total vulnerable functions, we extract the $m$ most similar matches to the query and determine the number of these (denoted as $m$) that are indeed vulnerable. 
Finally, we calculate the recall rate by $\frac{m}{k}$, which allows us to assess the effectiveness of different approaches.

\begin{figure}[t!]
  \centering
  \setlength{\abovecaptionskip}{2mm}
  \scalebox{1}{
  \includegraphics[width=0.8\linewidth]{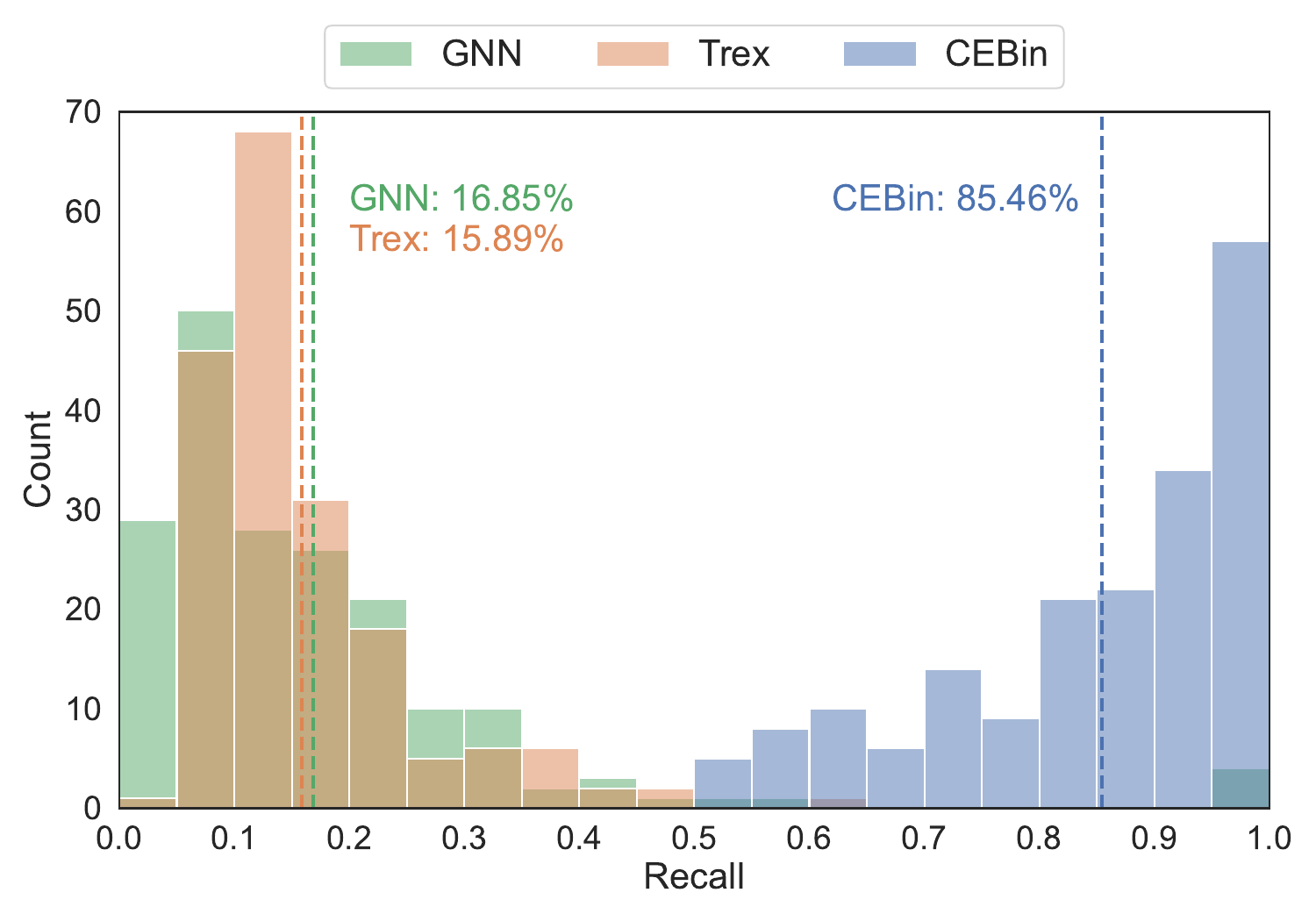}
  }
  \caption{The Vulnerability Search Results of \sysname and Trex. This figure represents the distribution of recall values of different methods. The dashed lines show the mean recall of different models.}
  \label{fig:Vulsearch-Results}
   \vspace{-1mm}
\end{figure}

\textbf{Result Analysis}
In Figure~\ref{fig:Vulsearch-Results}, we present the recall rate distribution for \sysname, GNN, and Trex on a vulnerability dataset. 
Among 187 CVEs, \sysname achieves an average recall rate of 86.46\%, while Trex and GNN have average recall of 15.89\% and 16.85\%, respectively. 
The recall rate distribution reveals that most of \sysname's recall rates fall within a range greater than 0.9. 
In contrast, the recall rates for Trex primarily fall within the 0.05 to 0.15 range. This observation implies that for large poolsize vulnerability searches, Trex and GNN experience a significant decrease in performance, while the impact on \sysname is relatively small. 
As a result, \sysname significantly outperforms both Trex and GNN. 
For instance, in CVE-2012-0884, CVE-2013-4353, CVE-2015-0288, and CVE-2015-1794, \sysname successfully identifies all vulnerabilities within poolsizes of over 60,000. 
Meanwhile, Trex and GNN could only detect less than 25\% of the vulnerable functions in these cases.

\iftrue
\textbf{Understanding Failure Cases}
We investigate various high-ranking failure cases in this experiment and identify several potential causes behind these inaccuracies:

\begin{packeditemize}
    \item \textbf{Truncation of excessively long functions.} Functions with tokens exceeding the maximum length are truncated when embedding. If two functions are similar before truncation, one might be misidentified as similar to the other after truncation. For example, in OpenSSL's CVE-2008-1672, the model confuses \texttt{ssl3\_send\_client\_key\_exchange} (vulnerable) with \texttt{dtls1\_send\_client\_key\_exchange} as they share highly similar semantic information within the maximum token length.

    \item \textbf{Insensitivity to specific instructions.} If functions only differ by very few instructions and the model cannot adequately distinguish them, it may consider them as similar. In OpenSSL's CVE-2012-2110, the model treats \texttt{CRYPTO\_realloc\_clean}~(vulnerable) and \texttt{CRYPTO\_realloc} as alike, despite the latter having three fewer consecutive function calls.

    \item \textbf{Function name altered during compilation.} Compiler may modify function names leading to misjudgments. In curl's CVE-2022-27780, the model identifies the vulnerable function as \texttt{hostname\_check.isra.1} instead of \texttt{hostname\_check}. After manually examining, we confirm that the misjudgment stems from the compiler changing the function name. 
    
\end{packeditemize}
\fi

\iftrue
\subsection{Generalization Performance~(RQ4)}\label{section:Generalization}
As previously discussed and shown in Table~\ref{tab:Trex-Poolsize-10000}, and Figure~\ref{fig:Trex-Recall@1-Poolsize}, after training on the Cisco Dataset, we observe that \sysname achieves excellent results on the Trex Dataset, demonstrating its strong generalization capability. 
To further investigate the generalization of \sysname, we conduct an even more challenging experiment. 
We label the models trained separately on BinaryCorp and Cisco as \sysname-BinaryCorp and \sysname-Cisco. 
Notably, BinaryCorp is a cross-optimization (XO) dataset compiled on x64 using one compiler, GCC 11.0, while the Cisco Dataset comprises six architectures and eight compilers. 
Because \sysname-BinaryCorp is trained based on cross-optimization tasks, we measure its performance in terms of cross-optimization BCSD on the Cisco Dataset and Trex Dataset for different architectures, alongside \sysname-Cisco.
Table~\ref{tab:ablation-xo-cisco} indicates that \sysname-BinaryCorp outperforms \sysname-Cisco in the XO task on x64 and other architectures, and on the Trex Dataset, with an average recall@1 increase of 0.028. The full results are available at Table~\ref{tab:cisco-xc}--\ref{tab:cisco-xo} in the Appendix.

The results are exciting and indicate that a model fine-tuned with XO dataset from one architecture can be readily transferred to other architectures. 
We believe that this outcome is ample evidence that the IL-based pre-trained \sysname model effectively normalizes binary functions across different architectures and extracts robust semantic features. 
It is worth noting that \sysname-BinaryCorp performs better than \sysname-Cisco, which we speculate may be attributed to the BinaryCorp dataset containing 1,612 projects and offering more diverse binary functions.
The Cisco Dataset consists of only seven projects, meaning that it may not be sufficiently diversified from a function semantics perspective, despite encompassing many different architectures and compiler variations. The use of a pre-trained model with IL has to some extent mitigated the effects of different architectures, and the BinaryCorp dataset is more diversified than the Cisco and Trex datasets, that is the reason \sysname-BinaryCorp performs better than \sysname-Cisco.

\begin{table}[t!]
\centering
\caption{Recall@1 comparison between \sysname-Cisco and \sysname-BinaryCorp of cross-optimization task on Cisco dataset for poolsize=10,000. (Simplified table)}
\scalebox{0.78}{
\begin{tabular}{c|cc|c}
\Xhline{1pt}
\multirow{2}{*}[-1.3ex]{\textbf{Architecture}} & \multicolumn{2}{c|}{\textbf{Model}} & \multirow{2}{*}[-1.3ex]{\textbf{Improvement}} \\ \cline{2-3}
 & \multicolumn{1}{c|}{\makecell[c]{\textbf{\sysname-}\\\textbf{Cisco}}} & \makecell[c]{\textbf{\sysname-}\\\textbf{BinaryCorp}} &  \\ \hline
x86 & \multicolumn{1}{c|}{0.984} & 0.988 & $\uparrow$ 0.004 \\ \hline
x64 & \multicolumn{1}{c|}{0.968} & 0.978 & $\uparrow$ 0.010 \\ \hline
MIPS32 & \multicolumn{1}{c|}{0.979} & 0.991 & $\uparrow$ 0.012 \\ \hline
MIPS64 & \multicolumn{1}{c|}{0.997} & 0.988 & $\downarrow$ 0.009 \\ \hline
ARM32 & \multicolumn{1}{c|}{0.962} & 0.967 & $\uparrow$ 0.005 \\ \hline
ARM64 & \multicolumn{1}{c|}{0.969} & 0.973 & $\uparrow$ 0.004 \\ \hline
\textbf{Average} & \multicolumn{1}{c|}{0.977} & \textbf{0.981} & \textbf{$\uparrow$ 0.004} \\ \Xhline{1pt}
\end{tabular}
}\label{tab:ablation-xo-cisco}
\end{table}


 

\begin{table}[t!]
\centering
\caption{Recall@1 comparison between \sysname-Cisco and \sysname-BinaryCorp of cross-optimization task on Trex dataset for poolsize=10,000.}
\scalebox{0.78}{
\begin{tabular}{c|cc|c}
\Xhline{1pt}
\multirow{2}{*}[-1.3ex]{\textbf{Architecture}} & \multicolumn{2}{c|}{\textbf{Model}} & \multirow{2}{*}[-1.3ex]{\textbf{Improvement}} \\ \cline{2-3}
 & \multicolumn{1}{c|}{\makecell[c]{\textbf{\sysname-}\\\textbf{Cisco}}} & \makecell[c]{\textbf{\sysname-}\\\textbf{BinaryCorp}} &  \\ \hline
x86 & \multicolumn{1}{c|}{0.907} & 0.948 & $\uparrow$ 0.041 \\ \hline
x64 & \multicolumn{1}{c|}{0.889} & 0.945 & $\uparrow$ 0.056 \\ \hline
MIPS32 & \multicolumn{1}{c|}{0.851} & 0.885 & $\uparrow$ 0.034 \\ \hline
MIPS64 & \multicolumn{1}{c|}{1.000} & 1.000 & $-$  \\ \hline
ARM32 & \multicolumn{1}{c|}{0.862} &  0.909 & $\uparrow$ 0.047 \\ \hline
ARM64 & \multicolumn{1}{c|}{0.925} & 0.917 & $\downarrow$ 0.008 \\ \hline
\textbf{Average} & \multicolumn{1}{c|}{0.906} & \textbf{0.934} & $\uparrow$ \textbf{0.028} \\ \Xhline{1pt}
\end{tabular}
\label{tab:ablation-xo-trex}
}
\end{table}

\fi

\subsection{Inference Cost~(RQ5)}\label{section:evaluation-execution-cost}

\begin{table}[t!]
\centering
\caption{Inference speed (seconds) of GNN, Trex, \sysname-E and \sysname on various poolsize.}
\scalebox{0.8}{
\begin{tabular}{ccccc}
\Xhline{1pt}
\multicolumn{1}{c}{}      & \multicolumn{4}{c}{\textbf{Poolsize}}          \\ \hline
\textbf{Model}                     & 100   & 10,000 & 1,000,000 & 4,000,000 \\ \hline
GNN                       & 0.004 & 0.004 & 0.012     & 0.037     \\ \hline
Trex                      & 0.018 & 0.018 & 0.050     & 0.182     \\ \hline
\sysname-E                   & 0.807    & 0.814 &  0.887 & 1.034        \\ \hline
\sysname & 2.717    & 2.772 &  2.898 & 3.105 \\   \Xhline{1pt}
\end{tabular}
\label{tab:inference-cost}
}
\end{table}
We evaluate the inference cost of \sysname against varying poolsizes and compare it to baselines. 
For embedding-based approaches, we pre-compute the corresponding vectors for the function pool. 
When assessing a new query function, we measure the cost from acquiring the embedding vector with the embedding-based model to comparing it with each function in the pool to obtain the results. 
The experimental results in Table~\ref{tab:inference-cost} reveal the time each method requires for handling different poolsizes.

Though \sysname's inference cost is relatively higher than the baseline, it does not increase rapidly as the poolsize expands. 
Even with a poolsize of 4 million, \sysname can process it in only 3.1 seconds, thereby demonstrating scalability for real-world software supply chain 1-day vulnerability discovery tasks.

\section{Limitations}
\sysname suffers from several limitations. First, while we have demonstrated the good performance of \sysname's BCSD on public datasets, we have only tested the real-world performance of the BCSD model in scenarios of 1-day vulnerability detection. Future work can conduct more comprehensive downstream tasks to address the challenges of applying BCSD technology in real-world senarios.

Second, there exist cloned functions across different projects in all datasets, thus leading to false negatives in the datasets. Even though the proportion is low, it might bring some negative impacts on the model performance and the evaluation of the final results. Future works can explore ways to construct a better dataset, such as performing more fine-grained deduplication to ensure the accuracy of experiments and enhance the performance of the model.

Third, our work only focuses on coarse-grained function level similarity detection. We cannot perfectly solve the 1-day vulnerability detection problem as the general BCSD can't distinguish whether a function is patched or not. Future works can combine BCSD solutions with fine-grained techniques such as directed fuzzing to better detect 1-day vulnerabilities.







\section{Conclusion}
In this paper, we propose \sysname, a novel cost-effective binary code similarity detection framework that bridges the gap between embedding-based and comparison-based approaches. \sysname employs a refined embedding-based approach to extract robust features from code, efficiently narrowing down the range of similar code. Following that, it uses a comparison-based approach to implement pairwise comparisons and capture complex relationships, significantly improving similarity detection accuracy. Through comprehensive experiments on three datasets, we demonstrate that \sysname outperforms state-of-the-art baselines in various scenarios, such as cross-architecture, cross-compiler, and cross-optimization tasks. We also showcase that \sysname successfully handles the challenge of large-scale function search in binary code similarity detection, making it an effective tool for real-world applications, such as detecting 1-day vulnerabilities in large-scale software ecosystems.





\iftrue
\newpage
\appendix

\section{Datasets} 
\label{sec:datasets}
We evaluate \sysname with the following three datasets. We label functions as equal if they share the same name and were compiled from the same source code.

\begin{packeditemize}
\item \textbf{BinaryCorp}~\cite{jtrans} is sourced from the ArchLinux official repositories and Arch User Repository. 
It contains tens of thousands of software, including editors, instant messengers, HTTP servers, web browsers, compilers, graphics libraries, and cryptographic libraries. 
Compiled using GCC 11.0 on X64 with various optimizations, it features a highly diverse set of projects.

\item \textbf{Cisco Dataset}~\cite{cisco} comprises seven popular open-source projects (ClamAV, Curl, Nmap, OpenSSL, Unrar, Z3, and zlib). It yields 24 distinct libraries upon compilation. 
Libraries are compiled using two compiler families, GCC and Clang, across four major versions each. 
Targeting six different ISAs (x86, x64, ARM32, ARM64, MIPS32, and MIPS64) and five optimization levels (O0, O1, O2, O3, and Os), the Cisco dataset enables cross-architecture evaluation and analysis of compiler versions, while containing a moderate project count.

\item \textbf{Trex Dataset}~\cite{pei2022learning} is built upon binaries released by~\cite{pei2022learning}, which consists of ten libraries chosen to avoid overlap with the Cisco dataset (binutils, coreutils, diffutils, findutils, GMP, ImageMagick, libmicrohttpd, libTomCrypt, PuTTY, and SQLite). 
Precompiled for x86, x64, ARM32, ARM64, MIPS32, and MIPS64, this dataset offers four optimization levels (O0, O1, O2, O3) and GCC-7.5. Similar to the Cisco Dataset, the Trex dataset facilitates cross-architecture and cross-optimization evaluation.
\end{packeditemize}

\section{Baselines}
\label{sec:baselines}

\begin{packeditemize} 
\item \textbf{Genius}~\cite{feng2016scalable}: is a non-deep learning approach extracting raw features as attributed control flow graphs and employing locality-sensitive hashing (LSH) to generate numeric vectors for vulnerability search.

\item \textbf{Gemini}~\cite{xu2017neural} extracts manually-crafted features for each basic block and then uses GNNs to learn the CFG representation of the target functions.

\item \textbf{SAFE}~\cite{massarelli2019safe} employs an RNN with attention mechanisms, this baseline generates a representation of the analyzed function using assembly instructions as input.

\item \textbf{Asm2Vec}~\cite{ding2019asm2vec} applies random walks on the CFG and uses the PV-DM model to jointly learn embeddings of the function and instruction tokens.

\item \textbf{GraphEmb}~\cite{massarelli2019investigating} learns embeddings of instruction tokens and uses Structure2vec~\cite{dai2016discriminative} to combine embeddings and generate the final representation.

\item \textbf{OrderMatters}~\cite{yu2020order}: concates two types of embeddings. It uses BERT to create an embedding for each basic block and
employs a CNN on the CFG to generate the second type of embedding.

\item \textbf{Trex}~\cite{pei2022learning} introduces a dynamic component extracting function traces based on a hierarchical transformer and micro-traces. This cross-architecture solution is built on the Transformer.

\item \textbf{GNN and GMN}~\cite{li2019graph} proposes graph-matching models for graph matching. Previous work~\cite{cisco} evaluates the two approaches on the Cisco and Trex datasets and the two approaches achieved the SOTA performance.

\item \textbf{jTrans}~\cite{jtrans} is a Transformer-based approach that embeds control flow information of binary code into Transformer-based language models with a jump-aware representation of binaries.

\end{packeditemize}

\section{Complete Experimental Results on Cisco Datasets}

\iftrue
\begin{table}[h!]
\centering
\caption{Recall@1 comparison between \sysname-Cisco and \sysname-BinaryCorp of cross-compiler task on Cisco dataset for poolsize=10,000. (Full table)}
\label{tab:cisco-xc}
\scalebox{0.8}{
\begin{tabular}{c|c|c|c|c}
\Xhline{1pt}
 \multirow{2}{*}[-1.3ex]{\textbf{ISA}} & \multirow{2}{*}[-1.3ex]{\textbf{Optimization}} & \multicolumn{2}{c|}{\textbf{Model}} & \multirow{2}{*}[-1.3ex]{\textbf{Improvement}} \\ \cline{3-4}
 &  & \multicolumn{1}{c|}{\makecell[c]{\textbf{CEBIN-}\\\textbf{Cisco}}} & \makecell[c]{\textbf{CEBIN-}\\\textbf{BinaryCorp}} &  \\ \hline
\multicolumn{1}{c|}{\multirow{5}{*}{x86}} & O0 & 0.988 & 0.994 & $\uparrow$ 0.006 \\ \cline{2-5} 
\multicolumn{1}{c|}{} & O1 & 0.993 & 0.986 & $\downarrow$ 0.007 \\ \cline{2-5} 
\multicolumn{1}{c|}{} & O2 & 0.986 & 0.993 & $\uparrow$ 0.007 \\ \cline{2-5} 
\multicolumn{1}{c|}{} & O3 & 0.990 & 1.000 & $\uparrow$ 0.010 \\ \cline{2-5} 
\multicolumn{1}{c|}{} & Os & 0.993 & 0.986 & $\downarrow$ 0.007 \\ \hline
\multicolumn{1}{c|}{\multirow{5}{*}{x64}} & O0 & 0.976 & 0.968 & $\downarrow$ 0.008 \\
\end{tabular}
}
\end{table}
\fi

\iftrue
\begin{table}[h!]
\centering
\caption{Results of Cross-Architecture on Trex Dataset}
\label{tab:my-table}
\scalebox{0.8}{
\begin{tabular}{c|c|c|c|c}
\Xhline{1pt}
\multirow{2}{*}[-1.3ex]{\textbf{Compiler}} & \multirow{2}{*}[-1.3ex]{\textbf{Optimization}} & \multicolumn{2}{c|}{\textbf{Model}} & \multirow{2}{*}[-1.3ex]{\textbf{Improvement}} \\ \cline{3-4}
 &  & \multicolumn{1}{c|}{\makecell[c]{\textbf{CEBIN-}\\\textbf{Cisco}}} & \makecell[c]{\textbf{CEBIN-}\\\textbf{BinaryCorp}} &  \\ \hline
\multicolumn{1}{c|}{\multirow{5}{*}{\texttt{GCC7.5}}} & O0 & 0.940 & 0.792 & $\downarrow$ 0.148 \\ \cline{2-5} 
\multicolumn{1}{c|}{} & O1 & 0.879 & 0.801 & $\downarrow$ 0.078 \\ \cline{2-5} 
\multicolumn{1}{c|}{} & O2 & 0.871 & 0.786 & $\downarrow$ 0.085 \\ \cline{2-5} 
\multicolumn{1}{c|}{} & O3 & 0.837 & 0.756 & $\downarrow$ 0.081 \\ \cline{2-5} 
\multicolumn{1}{c|}{} & Os & N/A & N/A & N/A \\ \hline
\multicolumn{2}{c|}{\textbf{Average}} & \textbf{0.882} & 0.784 & \textbf{$\downarrow$ 0.098} \\ \Xhline{1pt}
\end{tabular}
}
\end{table}
\fi

\begin{table}[t!]
\centering
\caption{Recall@1 comparison between \sysname-Cisco and \sysname-BinaryCorp of cross-optimization task on Cisco dataset for poolsize=10,000. (Full table)}
\label{tab:cisco-xo}
\scalebox{0.9}{
\begin{tabular}{c|c|c|c|c}
\Xhline{1pt}
\multirow{2}{*}[-1.3ex]{\textbf{ISA}} & \multirow{2}{*}[-1.3ex]{\textbf{Compiler}} & \multicolumn{2}{c|}{\textbf{Model}} & \multirow{2}{*}[-1.3ex]{\makecell[c]{\textbf{Improve-}\\\textbf{ment}}} \\ \cline{3-4}
 &  & \multicolumn{1}{c|}{\makecell[c]{\textbf{CEBIN-}\\\textbf{Cisco}}} & \makecell[c]{\textbf{CEBIN-}\\\textbf{BinaryCorp}} &  \\ \hline
\multicolumn{1}{c|}{\multirow{8}{*}{x86}} & \texttt{Clang3.5} & 0.989 & 0.989 & $-$ \\ \cline{2-5}  
\multicolumn{1}{c|}{} & \texttt{Clang5.0} & 0.985 & 0.985 & $-$ \\ \cline{2-5} 
\multicolumn{1}{c|}{} & \texttt{Clang7.0} & 0.986 & 0.986 & $-$ \\ \cline{2-5} 
\multicolumn{1}{c|}{} & \texttt{Clang9.0} & 0.985 & 0.982 & $\downarrow$ 0.003  \\ \cline{2-5} 
\multicolumn{1}{c|}{} & \texttt{GCC4.8} & 0.982 & 0.993 & $\uparrow$ 0.011  \\ \cline{2-5} 
\multicolumn{1}{c|}{} & \texttt{GCC5.0} & 0.986 & 0.989 & $\uparrow$ 0.003  \\
\end{tabular}
}
\end{table}

\clearpage
\newpage

\setlength{\tabcolsep}{2pt}
\tablefirsthead{\toprule CVE & Function & \makecell[r]{\#Vuln / Poolsize } \\ \midrule}
\tablehead{\toprule CVE & Function & \makecell[r]{\small \#Vuln / \small Poolsize } \\ \midrule}
\tabletail{\hline}

\tablecaption{Functions related to CVE, the number of vulnerable functions and search poolsize}
\begin{supertabular}{l p{3.67cm} r l }
\label{table:cve-search-functions-poolsize}
\small CVE-2004-0421 & \small \texttt{p}\hspace{0pt}\texttt{n}\hspace{0pt}\texttt{g}\hspace{0pt}\texttt{\_}\hspace{0pt}\texttt{f}\hspace{0pt}\texttt{o}\hspace{0pt}\texttt{r}\hspace{0pt}\texttt{m}\hspace{0pt}\texttt{a}\hspace{0pt}\texttt{t}\hspace{0pt}\texttt{\_}\hspace{0pt}\texttt{b}\hspace{0pt}\texttt{u}\hspace{0pt}\texttt{f}\hspace{0pt}\texttt{f}\hspace{0pt}\texttt{e}\hspace{0pt}\texttt{r} & \small 202 / 35707 \\
\small CVE-2006-2937 & \small \texttt{a}\hspace{0pt}\texttt{s}\hspace{0pt}\texttt{n}\hspace{0pt}\texttt{1}\hspace{0pt}\texttt{\_}\hspace{0pt}\texttt{d}\hspace{0pt}\texttt{2}\hspace{0pt}\texttt{i}\hspace{0pt}\texttt{\_}\hspace{0pt}\texttt{e}\hspace{0pt}\texttt{x}\hspace{0pt}\texttt{\_}\hspace{0pt}\texttt{p}\hspace{0pt}\texttt{r}\hspace{0pt}\texttt{i}\hspace{0pt}\texttt{m}\hspace{0pt}\texttt{i}\hspace{0pt}\texttt{t}\hspace{0pt}\texttt{i}\hspace{0pt}\texttt{v}\hspace{0pt}\texttt{e} & \small 4 / 13864 \\
\small CVE-2008-0891 & \small \texttt{s}\hspace{0pt}\texttt{s}\hspace{0pt}\texttt{l}\hspace{0pt}\texttt{\_}\hspace{0pt}\texttt{p}\hspace{0pt}\texttt{a}\hspace{0pt}\texttt{r}\hspace{0pt}\texttt{s}\hspace{0pt}\texttt{e}\hspace{0pt}\texttt{\_}\hspace{0pt}\texttt{c}\hspace{0pt}\texttt{l}\hspace{0pt}\texttt{i}\hspace{0pt}\texttt{e}\hspace{0pt}\texttt{n}\hspace{0pt}\texttt{t}\hspace{0pt}\texttt{h}\hspace{0pt}\texttt{e}\hspace{0pt}\texttt{l}\hspace{0pt}\texttt{l}\hspace{0pt}\texttt{o}\hspace{0pt}\texttt{\_}\hspace{0pt}\texttt{t}\hspace{0pt}\texttt{l}\hspace{0pt}\texttt{s}\hspace{0pt}\texttt{e}\hspace{0pt}\texttt{x}\hspace{0pt}\texttt{t} & \small 33 / 26950 \\
\end{supertabular}

\fi
\clearpage
\newpage
\bibliographystyle{plain}
\bibliography{main}

\end{document}